  \pdfoutput=1
  \documentclass[graphicx,amssymb,amsmath,enumerate,11pt]{report}
  \usepackage{fancyhdr}
  \fancyheadoffset[]{0.1cm}

  \usepackage{epsfig}
  \usepackage{axodraw}
  \usepackage{url}
  \newcommand\mchapter[2]{\chapter*{#1}
  \vskip -0.5cm \noindent {\it \LARGE #2}
  \addcontentsline{toc}{chapter}{#1\\{\normalsize\it #2}}}

\def\3he{$^3$He}
\def\4he{$^4$He}
\def\7li{$^7$Li}
\def\6li{$^6$Li}
\def\9be{$^9$Be}
\def\10b{$^10$B}
\def\11b{$^11$B}
\def\la{\mathrel{\mathpalette\fun <}}
\def\ga{\mathrel{\mathpalette\fun >}}
\def\fun#1#2{\lower3.6pt\vbox{\baselineskip0pt\lineskip.9pt
  \ialign{$\mathsurround=0pt#1\hfil##\hfil$\crcr#2\crcr\sim\crcr}}}
\def\beq#1\eeq{\begin{equation}#1\end{equation}}

\def\eg{{\it e.g.},}
\def\Yp{Y$_{\rm P}$}
\def\hii{H\thinspace{$\scriptstyle{\rm II}$}}
\def\hi{H\thinspace{$\scriptstyle{\rm I}$}}
\newcommand{\Deln}{\ensuremath{\Delta{\rm N}_\nu}}

\newcommand{\epm}{\ensuremath{e^{\pm}\;}}
\def\etal{{\it et al}}

\begin{document}      

 \rhead{\bfseries Neutrinos And Big Bang Nucleosynthesis}

 \mchapter{Neutrinos And Big Bang Nucleosynthesis}
 {Gary Steigman}
 \label{ch-19}

\vspace{0.5cm}

\begin{center}
{\it Center for Cosmology and Astro-Particle Physics, Department of Physics, Department of Astronomy, \\
The Ohio State University, 191 W.~Woodruff Ave., Columbus, 43210 OH, USA\footnote{email: steigman.1@asc.ohio-state.edu}}
\end{center}

\vspace{0.5cm}

\begin{center}
{\bf Abstract}
\end{center}
According to the standard models of particle physics and cosmology, there should be a background of cosmic neutrinos in the present Universe, similar to the cosmic microwave photon background.  The weakness of the weak interactions renders this neutrino background undetectable with current technology.  The cosmic neutrino background can, however, be probed indirectly through its cosmological effects on big bang nucleosynthesis (BBN) and the cosmic microwave background (CMB) radiation.  In this BBN review, focused on neutrinos and, more generally on dark radiation, the BBN constraints on the number of ``equivalent neutrinos" (dark radiation), on the baryon asymmetry (baryon density), and on a possible lepton asymmetry (neutrino degeneracy) are reviewed and updated.  The BBN constraints on dark radiation and on the baryon density following from considerations of the primordial abundances of deuterium and helium-4 are in excellent agreement with the complementary results from the CMB, providing a suggestive, but currently inconclusive, hint of the presence of dark radiation and, they constrain any lepton asymmetry.  For all the cases considered here there is a ``lithium problem": the BBN-predicted lithium abundance exceeds the observationally inferred primordial value by a factor of $\sim 3$.

\newpage

\section{Introduction}
\label{sec-19:intro}

According to the standard models of particle physics and cosmology, neutrinos (known and hypothesized) are produced, thermalized, and contribute to the total energy density in the early, hot, dense Universe, regulating the early Universe expansion rate.  Indeed, at the time of big bang nucleosynthesis (BBN), the contributions to the energy density from baryons, dark matter, and dark energy are all subdominant to those from the thermal populations of photons, electrons (\epm pairs), and neutrinos.  Since the abundances of the elements formed during the first few minutes of the evolution of the Universe depend on the competition between the universal expansion rate and the nuclear and weak interaction rates, the very good agreement between the BBN predictions and observations (see, \eg,~ref.~\cite{19-Steigman:2007,19-Steigman:2008,19-Steigman:2010,19-Nollett:2011} for reviews and further references) depends crucially on the early Universe thermalization of neutrinos and places restrictions on the presence of too many (or too few) of them or, of too much ``dark radiation".  At present, BBN and the cosmic microwave background (CMB) radiation provide the only probes of the cosmic neutrino background.  In addition to their contribution to the total energy density, electron neutrinos and antineutrinos play a special role in regulating the production of \4he, the second most abundance element in the Universe.  An excess of electron neutrinos over electron antineutrinos (lepton asymmetry;  neutrino degeneracy) or, vice versa, will change the neutron-to-proton ratio during BBN, modifying, mainly, the BBN-predicted primordial helium abundance.

BBN provides a window on the early evolution of the Universe and a probe of particle physics (neutrino physics and more) beyond the standard model (SM).  The primordial abundances of the elements produced in observationally accessible abundances by BBN (primarily D, \3he, \4he, \7li) depend on three fundamental parameters related to cosmology and particle physics: the baryon abundance (related to the Universal baryon asymmetry), the expansion rate of the Universe at BBN (a probe of dark radiation and the cosmic neutrino background) and, any neutrino degeneracy (lepton asymmetry).  

\subsection{Baryon Density Parameter}

The most obvious of these parameters is related to the abundance of the reactants, the baryons (nucleons).  Although the very early Universe may have begun symmetric between matter and antimatter ($n_{\rm B} = n_{\rm {\bar B}}$), long before BBN some yet to be determined mechanism involving the interplay between particle physics (violation of the conservation of baryon number, violation of C and CP symmetries) and cosmology (out of equilibrium evolution) led to a small but crucial local asymmetry between the amount of matter and antimatter in the Universe.  After nucleon-antinucleon annihilation, the excess (nucleons, by definition) survives ($n_{\rm B} - n_{\rm \bar{B}} \rightarrow n_{\rm B} \equiv n_{\rm N}$) and the number of nucleons in a comoving volume is preserved up to the present epoch (and far into the future as well).  Since the nuclear reaction rates depend on the nucleon density, which decreases as the Universe expands, it is convenient to normalize the nucleon density to the photon density.  After \epm annihilation, the ratio of the nucleon number density to the photon number density is unchanged as the Universe expands and cools\footnote{The number of nucleons in a comoving volume is conserved.   Entropy conservation guarantees that, after \epm annihilation, the number of photons in a comoving volume is also conserved.}.  BBN depends on the baryon density parameter $\eta_{10}$, defined by
\begin{equation}
  \label{eq-19:1}
  \eta_{10} \equiv 10^{10}\eta_{\rm B} \equiv 10^{10}(n_{\rm B}/n_{\gamma}).
\end{equation}
The present value ($t = t_{0}$, when the photon (CMB) temperature is $T_{0} = 2.725$\,K) of the baryon density is often measured by comparing the nucleon mass density to the critical mass density ($\Omega_{\rm B} \equiv (\rho_{\rm B}/\rho_{crit})_{0}$) and, the critical mass density depends on the present value of the Hubble parameter, the Hubble constant ($H_{0} \equiv 100\,h\,{\rm km\,s^{-1}\,Mpc^{-1}}$) \cite{19-Steigman:2006},
\begin{equation}
  \label{eq-19:2}
  \Omega_{\rm B}h^{2} = \eta_{10}/273.9.
\end{equation}

Predicting the baryon asymmetry of the Universe is one of the key challenges confronting the search for new physics beyond the standard model.  BBN constraints on $\eta_{\rm B}$ can help to identify potentially successful models of new physics.

\subsection{Expansion Rate Parameter}

The scale factor, $a = a(t)$, describes the evolution of the expansion of the Universe.  During the early evolution of the Universe the expansion rate, as measured by the Hubble parameter, $H \equiv (1/a)da/dt$, is determined by the total energy density which, during those epochs, is dominated by the contributions from massless or extremely relativistic particles, ``radiation" (R).
\begin{equation}
  \label{eq-19:3}
  H^{2} = 8\pi G\rho/3,
\end{equation}
where $G$ is Newton's gravitational constant and $\rho = \rho_{\rm R}$.  New physics may lead to $\rho_{\rm R} \rightarrow \rho'_{\rm R}$ (dark radiation) or, to a modification of the cosmology (general relativity) $G \rightarrow G'$, replacing the SM expansion rate with $H \rightarrow H' \equiv SH$.  The expansion rate factor, $S$, quantifies any departure from the standard models of particle physics and/or cosmology.

Prior to the start of BBN and prior to \epm annihilation (\eg~$m_{e} \la T \ll m_{\mu}$) the only relativistic SM particles present are the photons (with $g_{\gamma} = 2$ degrees of freedom or helicities), the \epm pairs ($g_{e} = 4$), and the N$_{\nu} = 3$, left-handed neutrinos and their right-handed antineutrinos ($g_{\nu} = 2$N$_{\nu}$), so that $\rho_{\rm R} = \rho_{\gamma} + \rho_{e} + \rho_{\nu}$.  The evolution of the Universe can be scaled out by comparing the total energy density to the energy density in the CMB photons.  Prior to \epm annihilation, $T_{\gamma} = T_{e} = T_{\nu}$, so that accounting for the different contributions to $\rho_{\rm R}$ from relativistic fermions and bosons,
\begin{equation}
  \label{eq-19:4}
  {\rho_{\rm R} \over \rho_{\gamma}} = 1 + {\rho_{e} \over \rho_{\gamma}} + {\rm N}_{\nu}\bigg({\rho_{\nu} \over \rho_{\gamma}}\bigg) = 1 + {7 \over 8}\bigg[\bigg({4 \over 2}\bigg) + \bigg({3\times 2 \over 2}\bigg)\bigg] = {43 \over 8},
\end{equation}
for N$_{\nu} = 3$.  The contribution from possible dark radiation (\eg~sterile neutrinos) may be expressed in terms of an equivalent number of SM neutrinos, \Deln~\cite{19-SSG:1977}.  At BBN, which begins prior to \epm annihilation, N$_{\nu}$ = 3 + \Deln.  In this case
\begin{equation}
  \label{eq-19:5}
  \rho'_{\rm R} \equiv \rho_{\rm R} + \Delta{\rm N}_{\nu}\rho_{\nu},
\end{equation}
or
\begin{equation}
  \label{eq-19:6}
  {\rho'_{\rm R} \over \rho_{\gamma}} = {43 \over 8} + {7 \over 8}\Delta{\rm N}_{\nu} = {43 \over 8}\bigg(1 + {7\Delta{\rm N}_{\nu} \over 43}\bigg).
\end{equation}
Allowing for dark radiation, the expansion rate factor, $S$, is directly related to \Deln,
\begin{equation}
  \label{eq-19:7}
  S \equiv {H' \over H} = \bigg({\rho'_{\rm R} \over \rho_{\rm R}}\bigg)^{1/2} = \bigg(1 + {7\Delta{\rm N}_{\nu} \over 43}\bigg)^{1/2}.
\end{equation}

It should be kept in mind that new physics ($S_{\rm BBN} \neq 1$) may manifest itself as $G_{\rm BBN} \neq G_{0}$ instead of $\Delta{\rm N}_{\nu} \neq 0$.  In this case, comparing $G_{\rm BBN}$ when $T \ga m_{e}$ to its present value,
\begin{equation}
  \label{eq-19:8}
  G_{\rm BBN}/G_{0} = S_{\rm BBN}^{2} = 1 + 0.163\Delta{\rm N}_{\nu}.
\end{equation}

After \epm annihilation the only relativistic SM particles present are the photons and the neutrinos.  The SM neutrinos decouple prior to \epm annihilation, when $T \sim 2-3$\,MeV, so that when the \epm pairs annihilate, the photons are heated relative to the neutrinos.  On the assumption that the neutrinos are fully decoupled at \epm annihilation, $T_{\nu}/T_{\gamma} = (4/11)^{1/3}$ and, for the SM (\Deln~= 0),
\begin{equation}
  \label{eq-19:9}
  {\rho_{\rm R} \over \rho_{\gamma}} = 1 + \bigg({\rho_{\nu} \over \rho_{\gamma}}\bigg) = 1 + {21 \over 8}\bigg({T_{\nu} \over T_{\gamma}}\bigg)^{4/3} = 1 + {21 \over 8}\bigg({4 \over 11}\bigg)^{4/3} = 1.681.
\end{equation}
However, in the presence of dark radiation or, ``equivalent neutrinos" (decoupled, with $T = T_{\nu} \neq T_{\gamma}$),
\begin{equation}
  \label{eq-19:10}
  S^{2} = {\rho'_{\rm R} \over \rho_{\rm R}} = 1 + \bigg({1 \over 1.681}\bigg){7 \over 8}\bigg({4 \over 11}\bigg)^{4/3}\Delta{\rm N}_{\nu} = 1 + 0.135\Delta{\rm N}_{\nu}.
\end{equation}
Since the SM neutrinos aren't fully decoupled at \epm annihilation, they do share some of the energy (entropy) when the \epm pairs annihilate \cite{19-Mangano:2005}. This has the effect of increasing the relative contribution of the neutrinos to the total radiation density so that after \epm annihilation, N$_{\nu} = 3 + \Delta{\rm N}_{\nu} \rightarrow$\,N$_{eff} = 3.046 + \Delta{\rm N}_{\nu}$.  As a result, later in the evolution of the Universe (\eg~at recombination), $\rho_{\rm R}/\rho_{\gamma} \rightarrow 1.692$ and $\rho'_{\rm R}/\rho_{\gamma} \rightarrow 1.692 + 0.227\Delta{\rm N}_{\nu}$, so that for $T \ll m_{e}$,
\begin{equation}
  \label{eq-19:11}
  S^{2} = {\rho'_{\rm R} \over \rho_{\rm R}} = 1 + \bigg({1 \over 1.692}\bigg){7 \over 8}\bigg({4 \over 11}\bigg)^{4/3}\Delta{\rm N}_{\nu} = 1 + 0.134\Delta{\rm N}_{\nu}.
\end{equation}
Of course, this post-BBN relation between the expansion rate ($S$) and the equivalent number of neutrinos (\Deln) is only relevant for those epochs when the Universe is radiation dominated.

BBN codes track the evolution of $S$ from $T \ga m_{e}$, prior to \epm annihilation, to $T \ll m_{e}$, well after \epm annihilation has ended.  Since it is important for BBN to follow the evolution of the neutron to proton ratio beginning when $T \ga$~few MeV, prior to \epm annihilation,
\begin{equation}
  \label{eq-19:12}
  S_{\rm BBN} \equiv (1 + 7\Delta{\rm N}_{\nu}/43)^{1/2} = (1 + 0.163\Delta{\rm N}_{\nu})^{1/2}.
\end{equation}
A BBN constraint  on $S$ is equivalent to one on \Deln~(or, on the ratio of $G_{\rm BBN}$ to its present value, $G_{0}$) and, later in the evolution of the Universe, N$_{eff} = 3.046 + \Delta{\rm N}_{\nu}$.  A BBN determination that $\Delta{\rm N}_{\nu}$ differs from zero at a significant level of confidence can provide evidence for new physics (dark radiation) such as the existence of one, or more, sterile neutrinos (thermally populated) or, a modification of the equations describing the expansion rate of the early Universe ($S_{\rm BBN} \neq 1$).

\subsection{Neutrino Degeneracy Parameter}

Since the charge neutrality of the Universe ensures that any electron excess is tied to the proton excess (the baryon asymmetry), a non-zero lepton asymmetry much larger than the baryon asymmetry ($\eta_{\rm B} \la 10^{-9}$) must be hidden in the neutrino sector.  An excess of neutrinos over antineutrinos (or, vice-versa) requires a non-zero neutrino chemical potential, $\mu_{\nu}$.  The dimensionless degeneracy parameter is the ratio of the neutrino chemical potential to the neutrino temperature, $\xi_{\nu} \equiv \mu_{\nu}/T_{\nu}$; $\xi_{\nu}$ is preserved as the Universe expands and cools.  In analogy with the parameterization of the baryon asymmetry by $\eta_{\rm B} \equiv (n_{\rm B} - n_{\rm {\bar B}})/n_{\gamma} \rightarrow n_{\rm B}/n_{\gamma}$, a lepton (neutrino) asymmetry may be parameterized by
\begin{equation}
  \label{eq-19:13}
\eta_{\rm L} = \eta_{\nu} = \Sigma_{\alpha}\,{(n_{\nu} - n_{\bar{\nu}})_{\alpha} \over n_{\gamma}} = {\pi^{3} \over 12\zeta(3)}\Sigma_{\alpha}\,\bigg[\bigg({\xi_{\alpha} \over \pi}\bigg) +  \bigg({\xi_{\alpha} \over \pi}\bigg)^{3}\bigg],
\end{equation}
where the sum is over the three SM neutrino flavors ($\alpha = e, \mu, \tau$).  Generally, mixing among the SM neutrinos ensures that the three chemical potentials are equilibrated.  In the following it is assumed that $\xi \equiv \xi_{e} = \xi_{\mu} = \xi_{\tau}$.  In this case,
\begin{equation}
  \label{eq-19:14}
\eta_{\rm L} = \eta_{\nu} = {\pi^{3} \over 4\zeta(3)}\bigg({\xi \over \pi}\bigg)\bigg[1 +  \bigg({\xi \over \pi}\bigg)^{2}\bigg].
\end{equation}

An asymmetry between electron neutrinos and electron antineutrinos has a direct effect on BBN through the charged current weak interactions which regulate the neutron-to-proton ratio ($p + e^{-} \leftrightarrow n + \nu_{e}, n + e^{+} \leftrightarrow p + \bar{\nu}_{e}, n \leftrightarrow p + e^{-} + \bar{\nu}_{e}$) (see, \eg~\cite{19-Beaudet:1976,19-Beaudet:1977,19-Boesgaard:1985,19-Kang:1992,19-Barger:2003,19-Kneller:2004,19-Simha:2008} and further references therein).  Since the relic abundance of \4he depends directly on the neutron-to-proton ratio when BBN begins (and during BBN), it provides a sensitive probe of any lepton asymmetry.  The abundances of the other light nuclides produced during BBN are less sensitive to $\xi$.

A subdominant effect (usually) of a non-negligible neutrino degeneracy ($\eta_{\rm L} \gg \eta_{\rm B}$) is to enhance to the contribution of the neutrinos to the early Universe energy density.  This is equivalent to a contribution to \Deln~where, for $\xi_{e} = \xi_{\mu} = \xi_{\tau} \equiv \xi$,
\begin{equation}
\label{eq-19:15}
\Delta{\rm N}_{\nu}(\xi) = {90 \over 7}\bigg({\xi \over \pi}\bigg)^{2}\bigg[1 + {1 \over 2}\bigg({\xi \over \pi}\bigg)^{2}\bigg].
\end{equation}
Note that for $|\xi| \la 0.1$, $\Delta{\rm N}_{\nu}(\xi) \la 0.013$, which is likely small compared with anticipated uncertainties in \Deln~inferred from BBN or the CMB.

At present and, likely for the foreseeable future, BBN provides the only window to a universal lepton asymmetry.

\section{The BBN Predicted Abundances}

For BBN within the context of the standard models of particle physics and cosmology (SBBN), along with some well defined extensions of them, only the light elements D, \3he, \4he, and \7li are produced in observationally interesting abundances.  The BBN-predicted relic abundances of these light nuclides depend on the three fundamental parameters introduced in \S\,\ref{sec-19:intro} \cite{19-Boesgaard:1985,19-Barger:2003,19-Barger:2003a,19-Kneller:2004,19-Steigman:2007,19-Simha:2008,19-Simha:2008a}.  Over limited, but interesting ranges of these parameters, the results for the abundances of these nuclides extracted from numerical BBN codes, are well fit (within the quoted errors) by \cite{19-Kneller:2004,19-Steigman:2007,19-Simha:2008,19-Simha:2008a},
\begin{equation}
\label{eq-19:d}
y_{\rm DP} \equiv 10^{5}({\rm D/H})_{\rm P} = 2.60(1 \pm 0.06)(6/\eta_{\rm D})^{1.6} = 45.7(1 \pm 0.06)\eta_{\rm D}^{-1.6},
\end{equation}
\begin{equation}
\label{eq-19:he}
{\rm Y_{\rm P}} = 0.2477 \pm 0.0006 + 0.0016(\eta_{\rm He} - 6) = 0.2381 \pm 0.0006 + 0.0016\eta_{\rm He},
\end{equation}
\begin{equation}
\label{eq-19:li}
y_{\rm LiP} \equiv 10^{10}({\rm Li/H})_{\rm P} = 4.82(1 \pm 0.10)(\eta_{\rm Li}/6)^{2}, \ \ {\rm A( Li}) \equiv 12 + {\rm log}({\rm Li/H})_{\rm P},
\end{equation}
where
\begin{equation}
\label{eq-19:etad}
\eta_{\rm D} = \eta_{10} - 6(S - 1) + 5\xi/4,
\end{equation}
\begin{equation}
\label{eq-19:etahe}
\eta_{\rm He} = \eta_{10} + 100(S - 1) - 575\xi/4,
\end{equation}
\begin{equation}
\label{eq-19:etali}
\eta_{\rm Li} = \eta_{10} - 3(S - 1) - 7\xi/4.
\end{equation}
The relation of $\eta_{\rm He}$ to $\xi$ in Eq.\,\ref{eq-19:etahe} is the one that appears in Kneller \& Steigman (2004)~\cite{19-Kneller:2004}.  An inadvertant typo in Steigman (2007)~\cite{19-Steigman:2007} was propagated in Simha \& Steigman (2008b)~\cite{19-Simha:2008a}.  The very small difference this typo generated, 575 vs. 574, has no effect on the quantitative results presented in those papers.  In the above equations, (D/H)$_{\rm P}$ and (Li/H)$_{\rm P}$ are the ratios by number of deuterium and of lithium (\7li) to hydrogen respectively, and \Yp~is the \4he mass fraction.  

There are some small but interesting changes in the numerical values in Equations \ref{eq-19:d}\,-\,\ref{eq-19:li} from earlier versions of these relations \cite{19-Kneller:2004,19-Steigman:2007,19-Simha:2008,19-Simha:2008a}; Equations \ref{eq-19:etad}\,-\,\ref{eq-19:etali} are unchanged.  In a recent paper, Nollett \& Holder (2011) \cite{19-Nollett:2011} called attention to the tension between the experimental and theoretical determinations of the $d(p,\gamma)^{3}{\rm He}$ cross section, important for predicting the BBN deuterium abundance.  Nollett \& Holder argue for preferring the theoretical calculation over the experimental result which, they suggest, may be affected by a normalization error.  Adoption of the theoretical calculation results in a $\sim 6\%$ reduction in the BBN-predicted D abundance.  I have preferred to ``split the difference", reducing the previously predicted abundance by 3\% but, doubling the error uncertainty (the error in the BBN-predicted value of D/H for a fixed value of $\eta_{\rm D}$) from 3\% to 6\%, resulting in the numerical values shown in Eq.\,\ref{eq-19:d}.

Given the role of the neutron-to-proton ratio at BBN on the predicted relic abundance of \4he, Y$_{\rm P}$ depends, albeit weakly, on the value of the neutron lifetime (mean life).  Quite recently, the Particle Data Group \cite{19-PDG:2011}, in response to discrepant experimental data, decided to change its recommended value for the neutron lifetime from $\tau_{n} = 885.7 \pm 0.8$\,s to $\tau_{n} = 881.5 \pm 1.5$\,s.  This change results in a small but noticeable reduction in the predicted value of Y$_{\rm P}$ by 0.0008 and, a very small increase in the associated uncertainty in the predicted value of \Yp, $0.0005 \rightarrow 0.0006$.  Also incorporated into Eq.\,\ref{eq-19:he} is the Mangano \etal~\cite{19-Mangano:2005} correction to the helium abundance resulting from the incomplete decoupling of the neutrinos at \epm annihilation.

The intense interest in recent years in the ``lithium problem(s)" has led to an extensive reevaluation of the relevant nuclear reaction rates \cite{19-CFO:2003,19-Cyburt:2004,19-CFO:2008}, leading to an increase in the BBN-predicted abundance by $\sim 12\%$ from the result presented in Steigman 2007~\cite{19-Steigman:2007}, further exacerbating the lithium problem to be discussed below.  This is reflected in Eq.\,\ref{eq-19:li}.  The $d(p,\gamma)^{3}{\rm He}$ rate which plays a role in the primordial deuterium abundance also impacts the BBN-predicted lithium abundance \cite{19-Nollett:2011} (K. Nollett, Private Communication).  This is taken into account in our error estimate.

It should be emphasized that these fits are {\bf not} analytic approximations to the results from a numerical BBN code.  Rather, they are fits to the results from such a code which are primarily {\bf simple}, involving a minimal number of numerical values, chosen to only one or two significant figures.  For limited but interesting ranges of the parameters ($5.5 \la \eta_{10} \la 6.5$, $0.85 \la S \la 1.15$ ($-1.7 \la \Delta{\rm N}_{\nu} \la 2.0$), $-0.1 \la \xi \la 0.1$) these fits agree with the numerical results from this and other codes within the quoted uncertainties.  

It is worth noting that the BBN-predicted abundances of D and \7li are mainly sensitive to the baryon abundance while that of \4he is more sensitive to non-standard physics (\Deln~and/or $\xi$).  However, both D and \7li are weakly dependent on \Deln~and/or $\xi$ and, \4he is weakly dependent on $\eta_{\rm B}$.

\section{The Observationally Inferred Primordial Abundances}
\label{sec-19:obsabund} 

Any conclusions about new physics (\eg~$\Delta{\rm N}_{\nu} \neq 0$?, $\xi \neq 0$?) based on BBN depend on the adopted primordial abundances.  The relic abundances of D and \4he are key to the conclusions reviewed and updated here.  It may have been noticed that \3he failed to be included in the discussion in the previous section.  The reason is that \3he is only observed in the interstellar medium of our galaxy, which consists of gas that has been processed through many generations of stars.  The large and uncertain corrections for post-BBN stellar processing make it difficult to infer the primordial abundance of \3he using the current data (see, \eg~\cite{19-Bania:2002} for discussion and further references).  In addition, the observationally inferred abundances of lithium in the oldest, most metal poor stars in the Galaxy are systematically lower, by factors of $\sim 3 - 4$, than the BBN-predicted values (the lithium problem or, one of several lithium problems).  Whether this discrepancy results from poorly understood corrections for stellar structure and/or evolution or, is a hint of new physics, remains unclear at present.  In the confrontation of the BBN predictions with the observational data only D and \4he will be used to constrain various combinations of $\eta_{10}, \Delta{\rm N}_{\nu}, \xi$, and the results will be used to {\bf predict} the primordial lithium abundance which will then be compared to the observations.  For a much more detailed, albeit not entirely up to date, discussion of the observational data and, in particular, the problems (real and potential) associated with them, see my recent reviews \cite{19-Steigman:2007,19-Steigman:2010}.
 
\begin{figure}[h!!]
\begin{center}
\vspace{1cm}
 \includegraphics*[scale=0.375]{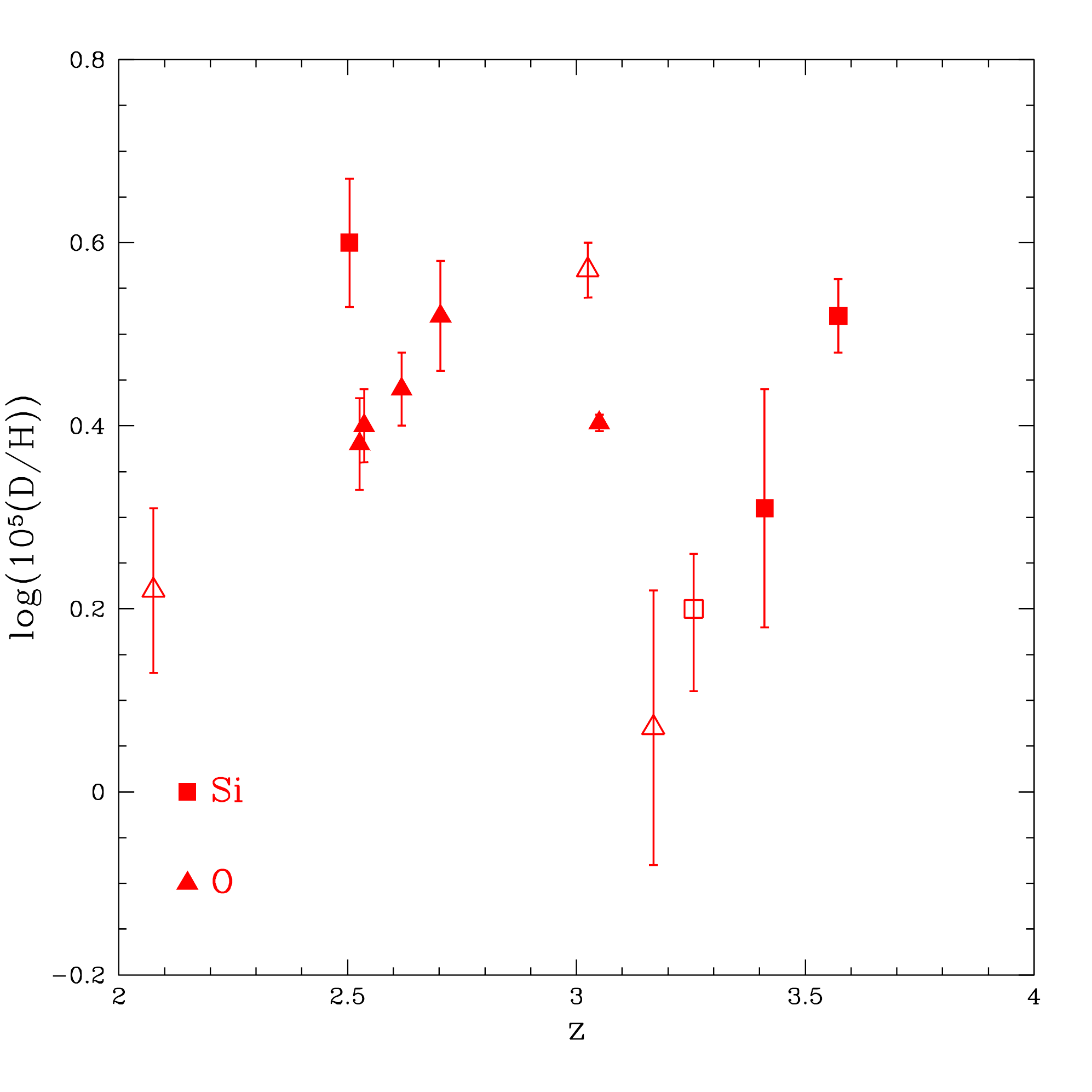}
\hspace{.5cm}
 \includegraphics*[scale=0.375]{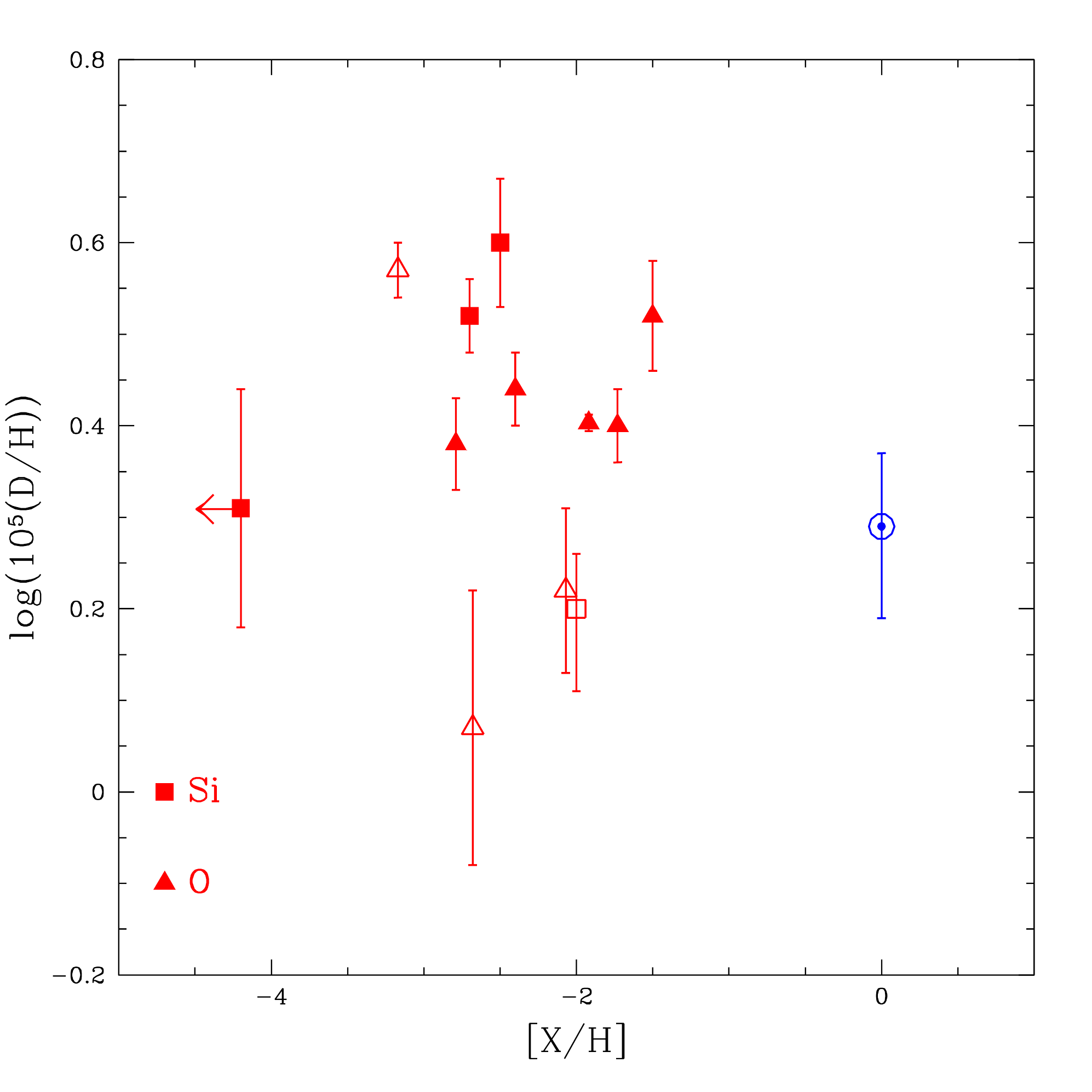}
\hspace{.5cm}
 \includegraphics*[scale=0.375]{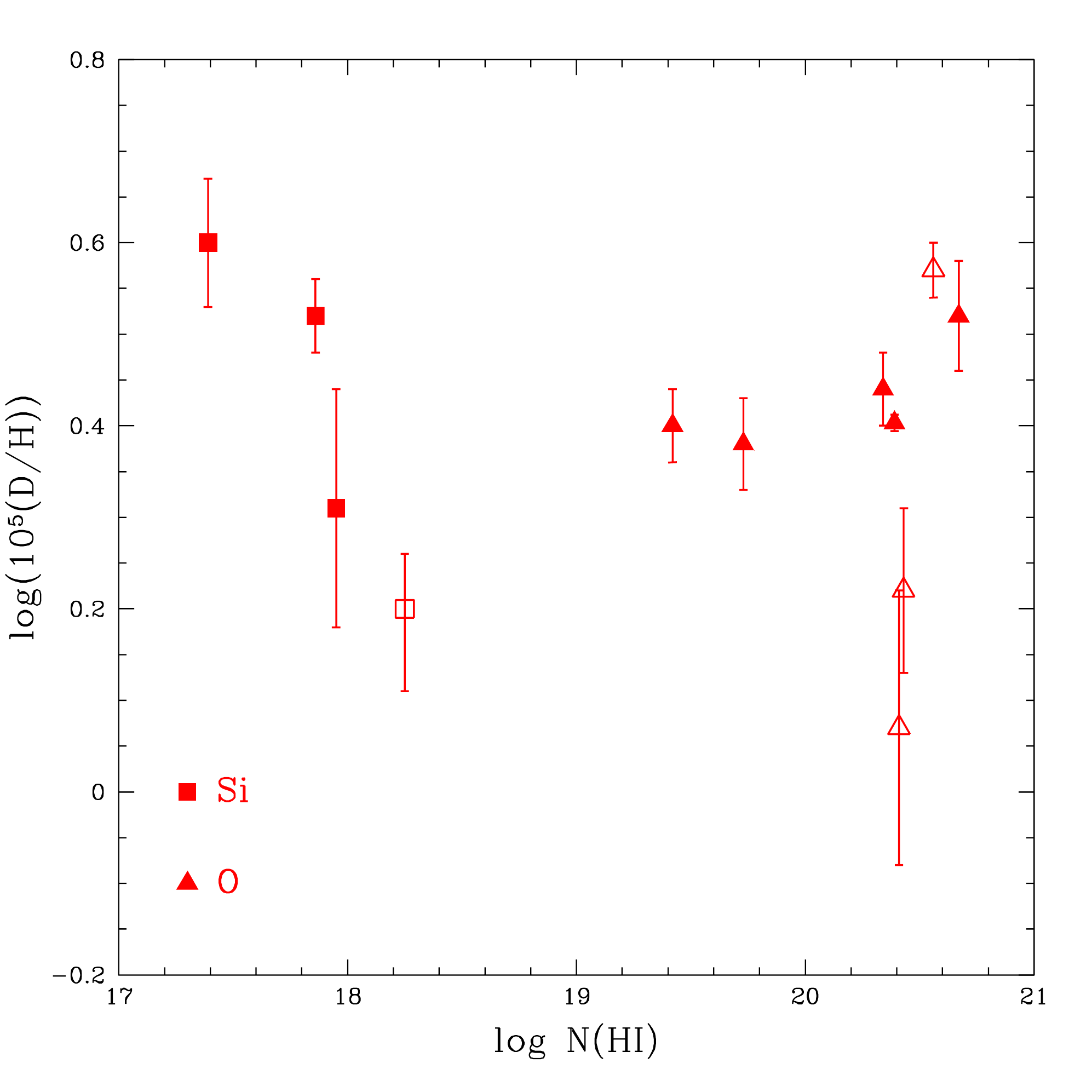}
\caption{The log of the deuterium abundances ($y_{\rm D} \equiv 10^{5}({\rm D/H})$), and their $1\,\sigma$ uncertainties, inferred from 12 low$-Z$, high$-z$ QSOALS.  In the upper left panel $y_{\rm D}$ is shown as a function of the redshift.  In the upper right panel $y_{\rm D}$ is shown as a function of the metallicity (squares for silicon and triangles for oxygen).  In the lower panel $y_{\rm D}$ is shown as a function of the neutral hydrogen column density.  The filled symbols reflect a subjective determination of the more robust determinations of D/H compared to the open symbols.  The blue solar symbol in the upper right panel is the solar system (pre-solar nebula) D abundance~\cite{19-GG:1998}.
  \label{fig-19:dobs}}
\end{center}
\end{figure}

\subsection{Primordial Deuterium}
\label{sec-19:primdeut} 

In the post-BBN Universe, as gas is cycled through successive generations of stars, deuterium is destroyed, not produced \cite{19-ELS:1976}.  The post-BBN evolution of deuterium is simple and monotonic (decreasing abundance).  As a result, observations of deuterium at high redshifts ($z$) and/or of gas at low metallicity ($Z$), where very little of the primordial gas has been cycled through stars, should provide a view of very nearly primordial deuterium.  While interesting on their own, observations of deuterium in the chemically evolved Galaxy or the solar system are of relatively little use in constraining the relic deuterium abundance.  Observations of D at high $z$ and low $Z$ are provided by the QSO absorption-line systems (QSOALS) \cite{19-Kirkman:2003,19-O'Meara:2006,19-Pettini:2008,19-Fumagalli:2011,19-Pettini:2012}.  Since deuterium is observed by absorption of background QSO light in the wings of the much larger hydrogen absorption, exquisite velocity information about the absorbing gas is crucial to a meaningful determination of the D/H ratio.  This, and other contributors to potential systematic errors, has limited the number of ``robust" D abundance determinations from high$-z$, low$-Z$, QSOALS.  In Fig.\,\ref{fig-19:dobs} are shown 12 high$-z$, low$-Z$ D/H determinations as a function of the absorbing redshift (upper left panel), of the metallicity (upper right panel), and of the \hi~column density (lower panel).  The open symbols reflect a subjective judgment of abundance determinations which may be more uncertain than indicated by their error bars (perhaps all the data should be plotted with open symbols).

For all 12 D abundance determinations, the weighted mean abundance is $<{\rm log}\,y_{\rm D}>\, = 0.42$.  However, it is clear (by eye) that there is an excessively large dispersion among the individual abundance determinations (\eg~the reduced $\chi^{2}$ for 11 degrees of freedom is $\chi^{2}/dof = 4$).  It is also clear from the three panels in Fig.\,\ref{fig-19:dobs} that the spread in abundances does not  correlate with either redshift, metallicity or, \hi~column density.  The absence of any correlations suggests that it is unlikely that the spread in the observed deuterium abundances results from post-BBN evolution.  Lacking a well motivated understanding of the cause(s) of the observed dispersion, various statistical approaches have been adopted for estimating the uncertainty when identifying the weighted mean abundance from the observations with the primordial deuterium abundance.  Consistent with these more sophisticated approaches, if the quoted errors for each of the data points is simply doubled (lowering the reduced $\chi^{2}/dof$ from 4 to 1), it leads to the following estimate of the relic abundance,
\begin{equation}
\label{eq-19:dabund}
{\rm log}\,y_{\rm DP} = 0.42 \pm 0.02 \ \ (y_{\rm DP} \equiv 10^{5}({\rm D/H})_{\rm P} = 2.63 \pm 0.12).
\end{equation}
This result is consistent with that quoted in Pettini \& Cooke (2012)~\cite{19-Pettini:2012}, who used a slightly different set of D/H observations\footnote{For their new, most precise individual deuterium abundance determination, Pettini \& Cooke \cite{19-Pettini:2012} find $y_{\rm D} = 2.53 \pm 0.05$ which, if identified with the primordial deuterium abundance, corresponds to $\eta_{\rm D} = 6.10 \pm 0.24$.}.  Adopting this estimate for the primordial D abundance, Eq.\,\ref{eq-19:d} results in
\begin{equation}
\label{eq-19:etadobs}
\eta_{\rm D} = 5.96 \pm 0.28.
\end{equation}

\subsection{Primordial Helium}
\label{sec-19:primhe}

\begin{figure}[h!!]
\begin{center}
\vspace{1cm}
\includegraphics*[scale=0.5]{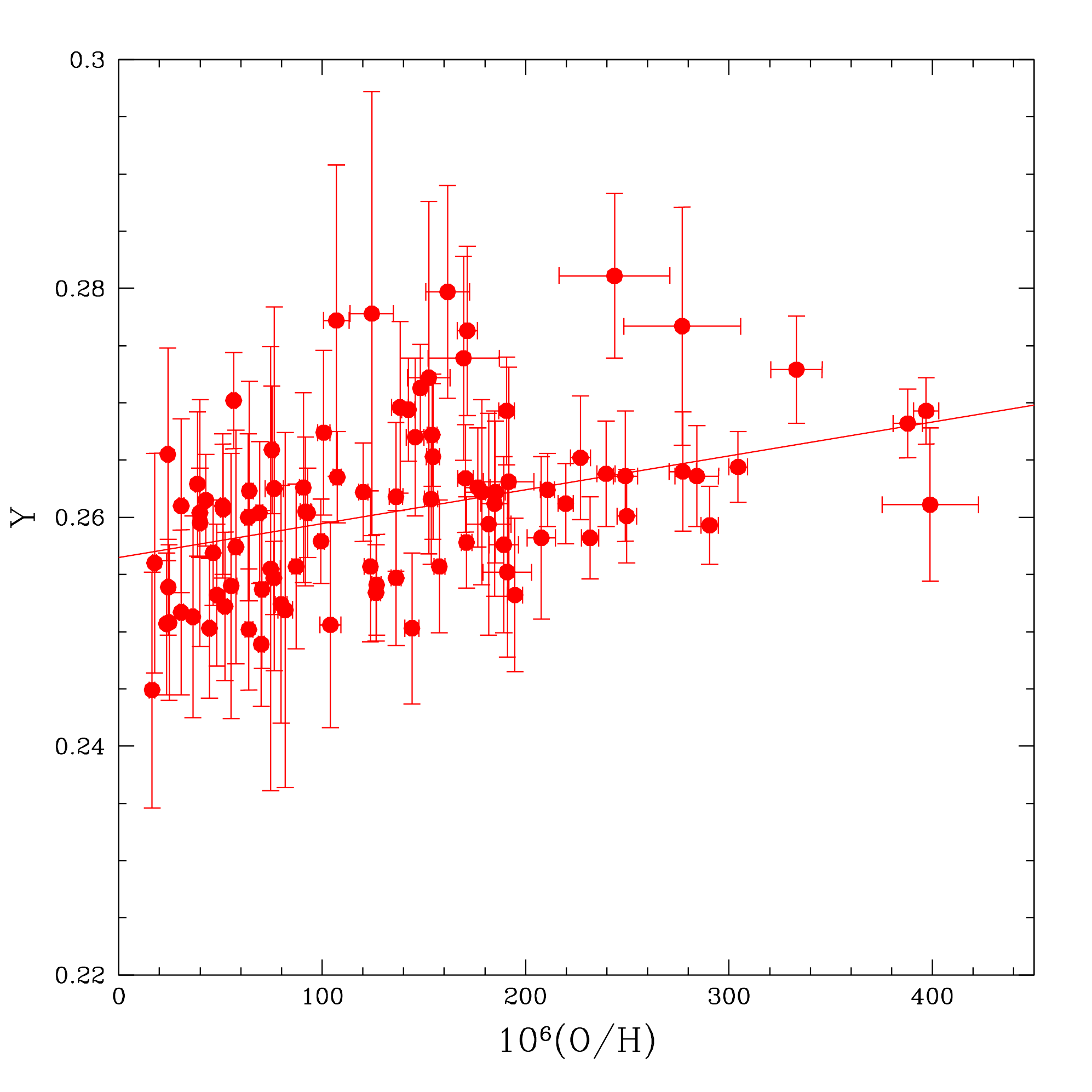}
\caption{Helium abundance (mass fractions, Y) determinations from the sample of extragalactic \hii~regions studied by Izotov \& Thuan (2010) \cite{19-IT:2010} as a function of the corresponding oxygen abundances (O/H by number).  The solid line is the Izotov \& Thuan best fit to a linear Y versus O/H correlation.
\label{fig-19:it}}
\end{center}
\end{figure}

As was the case for deuterium, the post-BBN evolution of helium (\4he) is simple and monotonic.  As gas is cycled through stars, hydrogen is burned to helium (and beyond) and the helium abundance increases with time and with metallicity.  The strategy, therefore, is to concentrate on determining the helium abundance in the most nearly primordial regions of low metallicity which, as with deuterium, lie outside of the Galaxy.  In \hii~regions, regions of hot, ionized gas, recombinations of hydrogen and helium result in observable emission lines which can be used, along with models of the \hii~regions and a knowledge of the associated atomic physics, to infer the helium abundance.  The current inventory of helium abundance determinations from relatively low metallicity, extragalactic \hii~regions approaches $\sim 100$ (Izotov \& Thuan 2010 (IT)~\cite{19-IT:2010}).  These data for the inferred helium mass fraction (Y) as a function of the corresponding oxygen abundance (O/H by number) are shown in Fig.\,\ref{fig-19:it}.  The sheer size of this data set leads to relatively small, formal statistical errors, magnifying the importance of taking proper account of the many possible sources of systematic errors.  While some have employed {\it a posteriori} selected subsets of the IT data for more detailed analyses, the sources and magnitudes of systematic errors have rarely been addressed.  As a result, the uncertainty in the inferred primordial helium mass fraction is currently dominated by systematic errors (both the known unknowns and the unknown unknowns).  As a result, estimating the size of the true uncertainty in the observational determination of \Yp~is largely guesswork.  From a linear fit to their Y -- O/H data, IT find the intercept, providing an estimate of the primordial helium abundance,
\begin{equation}
\label{fig-19:itY}
{\rm Y}_{\rm P} = 0.2565 \pm 0.0010\,(stat) \pm 0.0050\,(syst).
\end{equation}

In the analysis presented here the statistical and systematic errors are, arbitrarily, combined {\it linearly}, leading to the estimate of the primordial abundance adopted here,
\begin{equation}
\label{fig-19:Y}
{\rm Y}_{\rm P} = 0.2565 \pm 0.0060.
\end{equation}
This estimate of the relic \4he abundance is consistent with other, recent estimates based on analyses involving limited subsets of the IT data.  In some of those other analyses, a linear Y versus O/H fit is forced on data which is consistent with no correlation between Y and O/H.  Not surprisingly, the result of such analyses is a slope which is consistent with zero at less than $1\,\sigma$.  However, the large uncertainty in the slope inferred from such fits leads to an estimate of the intercept (\Yp) with excessively large errors, which have nothing to do with either the statistical or systematic errors.  The errors simply reflect the uncertainty in the slope for uncorrelated data.  Even worse, since these fits are consistent with an unphysical, negative Y -- O/H slope at $\la 1\,\sigma$, they lead to an unphysical upper bound to \Yp.

In combination with Eq.\,\ref{eq-19:he}, the relic abundance adopted here results in,
\begin{equation}
\label{eq-19:etaheobs}
\eta_{\rm He} = 11.50 \pm 3.77.
\end{equation}

\subsection{Primordial Lithium}

Compared to the post-BBN evolution of D and \4he, the evolution of lithium is more complicated and uncertain, similar to that of \3he.  As gas is cycled through stars, most of the pre-stellar lithium is destroyed.  However, some lithium may avoid nuclear burning if it remains in the cooler, surface regions of the coolest, lowest mass stars.  Observations suggest that some stars (the ``super-lithium rich" red giants), during some part of their evolution, are net producers of lithium, although it is not entirely clear if such stellar produced lithium is returned to the interstellar gas before being destroyed.  Finally, it is well known that collisions in the interstellar medium between cosmic rays, mainly alpha particles, and interstellar gas nuclei, primarily CNO nuclei, break up those nuclei producing lithium (\7li, along with \6li and, isotopes of Be and B).  The net effect of post-BBN production, destruction, and survival is difficult to model precisely.  However, there is observational evidence supporting a lithium abundance which increases along with the heavy element abundance (metallicity), suggesting an overall increase of the lithium abundance with time.

As we are interested in samples of the most nearly primordial material, the best (only) targets for determining the relic lithium abundance are the oldest, most metal poor stars in the Galaxy.  If the metallicity is sufficiently small, so that the material in these stars has suffered  very little processing, observations should find a ``lithium plateau".  That is, for such metal poor stars the lithium abundance should be uncorrelated with metallicity, revealing the primordial lithium abundance.  Since the observationally inferred lithium abundance (relative to hydrogen) is so small (by number, Li/H $\sim 10^{-10} - 10^{-9}$), it is common to measure it on a logarithmic scale by the quantity A(Li)~$\equiv 12 + {\rm log}({\rm Li/H}$).  The metallicity is usually quantified by comparing, also on a logarithmic scale, the iron abundance (Fe/H) to that in the Sun: [Fe/H]~$\equiv {\rm log (Fe/H) - log (Fe/H)}_{\odot}$.  Observations of stars with $-2.5 \la [{\rm Fe/H}] \la -1.0$ (those with $\sim 0.3\%$ to $\sim 1\%$ of the metallicity of the Sun) do appear to lie on a plateau, the ``Spite plateau" \cite{19-SS:1982}, at a level of A(Li)~$\approx 2.2 \pm 0.1$~\footnote{Recently, Nissen \& Schuster \cite{19-NS:2012} have suggested that stars with $-1.5 \la [{\rm Fe/H}] \la -0.7$ were formed with a lithium abundance close to primordial, which they estimate as A(Li)~$= 2.58 \pm 0.04\,(stat) \pm 0.10\,(syst)$, much closer to the BBN-predicted abundances discussed below.  They attribute the lower observed abundances to lithium depletion in the stellar atmospheres.}.  As will be seen below, all of the BBN predictions are close to A(Li)~$\approx 2.7$ or, a factor of $\sim 3$ higher than suggested by the observations \cite{19-Asplund:2006,19-Aoki:2009,19-Lind:2009}.  This is one of the lithium problems.  However, as recent observations of even more metal poor stars ($-3.5 \la {\rm [Fe/H]} \la -2.5$) have accumulated, the lithium plateau appears to be transforming into a ``lithium cliff", with lower lithium abundances (A(Li)~$\approx 2.1 \pm 0.1$) correlating with lower metallicities \cite{19-Asplund:2006,19-Aoki:2009,19-Lind:2009}.  This trend is puzzling and not understood at present; another lithium problem.  Since this second lithium problem is currently unresolved, it is not clear if the lower abundances are too be interpreted as suggesting an even lower value for the primordial lithium abundance, further exacerbating the original lithium problem or, if they are telling us something about the evolution of the oldest, most metal poor stars in the Galaxy which will require us to reevaluate both lithium problems, along with the observationally inferred value of the primordial lithium abundance.

For these reasons (the lithium problems), lithium does not provide a useful probe of BBN at present.  For a recent review of the current lithium data and possible resolution of the lithium problem(s), see Fields 2011~\cite{19-Fields:2011}.  In the following, BBN, possibly in combination with the CMB, will constrain the key parameters using the observationally inferred primordial D and \4he abundances and, those parameter combinations will be used to predict the BBN abundance of lithium (reinforcing the problem of the low observed abundances).

\section{BBN Constraints On The Fundamental Parameters}
\label{sec-19:constraints}

The BBN-predicted primordial abundances depend on all three parameters: $\eta_{10}, \Delta{\rm N}_{\nu}, \xi$.  Because of the uncertainty of if, or how, to use lithium (\7li), here we limit ourselves to employing only the D and \4he abundances.  Then, without recourse to additional, non-BBN data, only two of these parameters can be constrained at a time.  Here, we will consider constraints  on the baryon density ($\eta_{10}$) from BBN (using D as our primary baryometer) and on the parameter pairs \{$\eta_{10}, \Delta{\rm N}_{\nu}$\} and \{$\eta_{10}, \xi$\} (using D and \4he in combination) and, we will comment on the result of using complementary data on $\eta_{10}$ or \Deln~from the CMB in order to constrain all three parameters simultaneously.

\subsection{BBN Constraint On The Standard Model Baryon Density ($\Delta{\rm N}_{\nu} = 0 = \xi$)}
\label{sec-19:sm}

Before entertaining the possibility of new physics (dark radiation and/or lepton asymmetry), the D abundance may be used to provide a standard model BBN (SBBN) constraint on the baryon density; the SBBN-predicted D abundance is much more sensitive to the baryon density than is the helium abundance.  Assuming that $\Delta{\rm N}_{\nu} = 0 = \xi$, $\eta_{\rm D} = \eta_{10} = 5.96 \pm 0.28$ ($\Omega_{\rm B}h^{2} = 0.0218 \pm 0.0010$), in excellent agreement with the value found from the CMB (\eg~from WMAP~\cite{19-Komatsu:2011}).

This SBBN-inferred value of the baryon density may be used to predict the primordial abundances of the other light nuclides, \3he, \4he and \7li.  There is good agreement between the SBBN-predicted value of the \3he abundance and an upper bound to it inferred from observations stellar-processed gas in the Galaxy \cite{19-Bania:2002}.  For SBBN the relic \4he~mass fraction is predicted to be Y$_{\rm P} = 0.2476 \pm 0.0007$, a value smaller than the observationally-inferred abundance adopted here but, within $\sim 1.5\,\sigma$ of it (see \S\,\ref{sec-19:primhe}).    The problem for SBBN is lithium.  For the above value of the baryon density and for $\Delta{\rm N}_{\nu} = 0 = \xi$, the predicted primordial lithium abundance is A(Li) $= 2.68 \pm 0.06$.  This is higher, by a factor of $\sim 3-4$, than the values inferred from observations of the most metal poor (most nearly primordial) stars in the Galaxy.  This is (one of) the lithium problem(s).
 
\subsection{BBN Constraints On The Baryon Density ($\eta_{10}$) And Dark Radiation (\Deln)}
\label{sec-19:etannu}

\begin{figure}[h!!]
\begin{center}
\vspace{1cm}
\includegraphics*[scale=0.5]{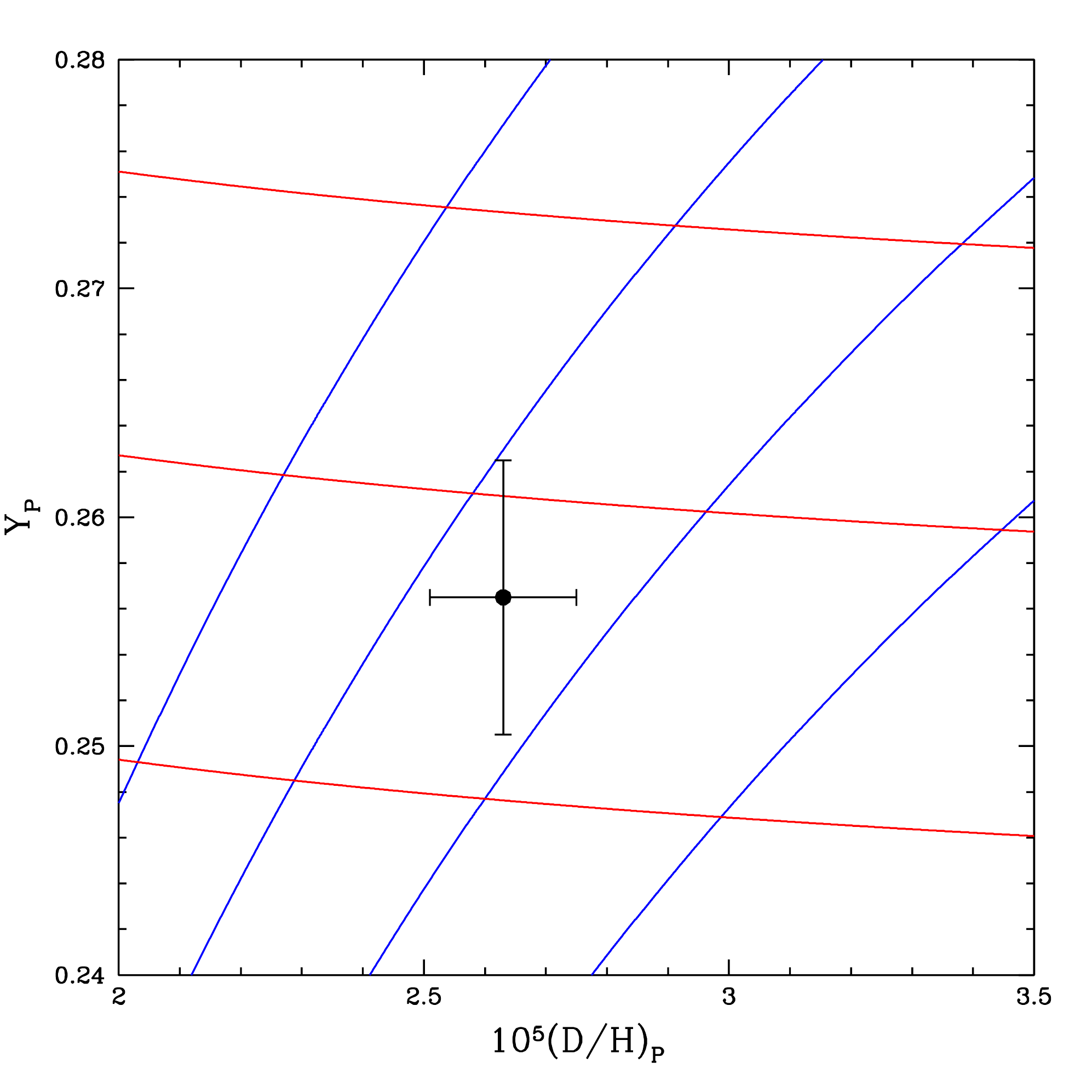}
\caption{Contours of constant values of \Deln~(red) and $\eta_{10}$ (blue) in the Y$_{\rm P} - y_{\rm DP}$ plane.  From bottom to top the red curves correspond to $\Delta{\rm N}_{\nu} = 0, 1, 2$.  From left to right the blue curves correspond to $\eta_{10} = 7.0, 6.5, 6.0, 5.5$.  Also shown (filled circle and error bars) are the adopted primordial abundances of D and \4he and their $1\,\sigma$ uncertainties.
  \label{fig-19:hevsd}}
\end{center}
\end{figure}

\begin{figure}[h!!]
\begin{center}
\vspace{1cm}
 \includegraphics*[scale=0.5]{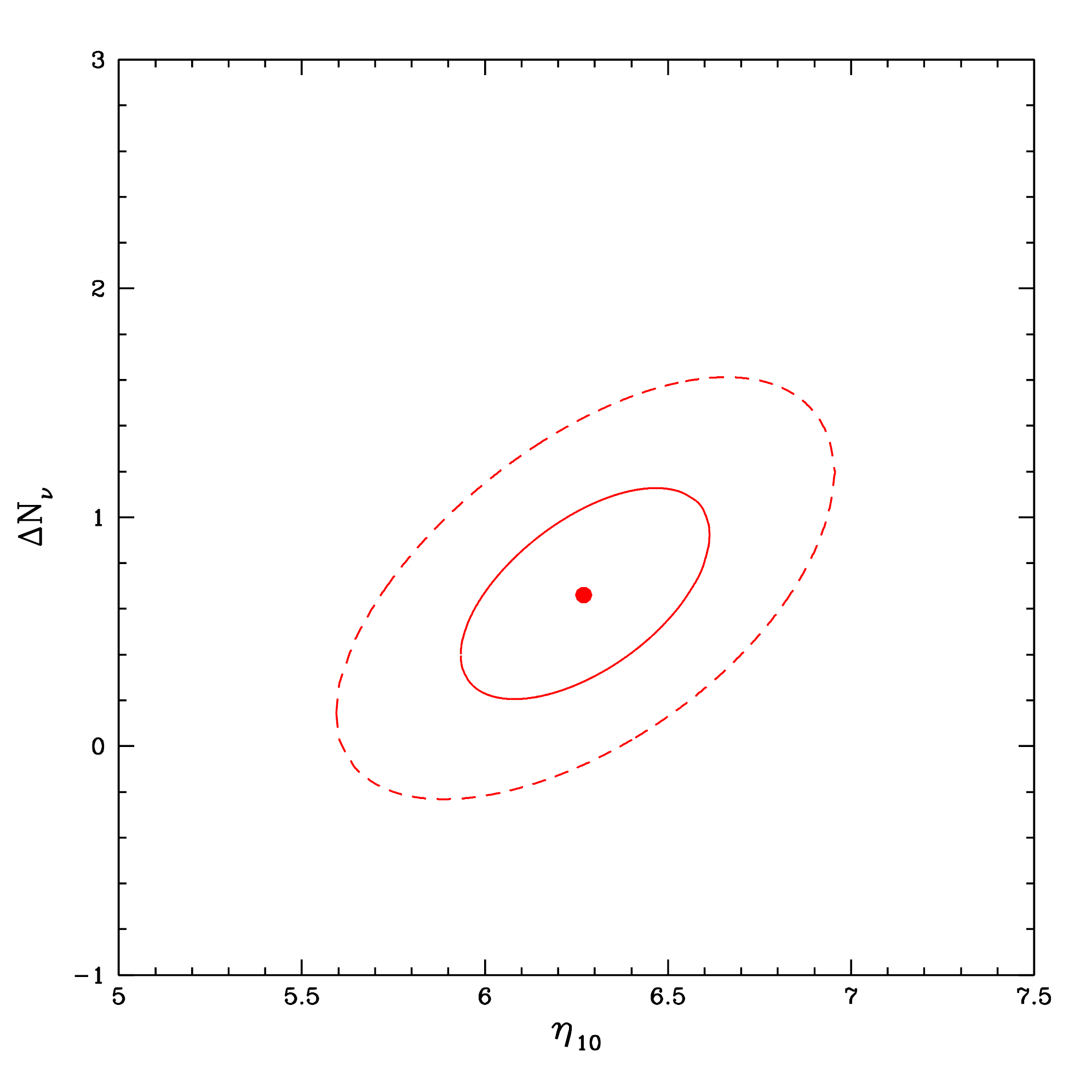}
\caption{The BBN-inferred 68\% (solid) and 95\% (dashed) contours in the \Deln~-- $\eta_{10}$~plane derived from D and \4he assuming that $\xi = 0$.
\label{fig-19:nnuvseta95}}
\end{center}
\end{figure}

If it is assumed that there is no lepton asymmetry, $\xi = 0$, Eqs.\,\ref{eq-19:etad} \& \ref{eq-19:etahe} may be solved for $\eta_{10}$ and $S$ (\Deln) in terms of $\eta_{\rm D}$ and $\eta_{\rm He}$.  
\begin{equation}
\label{eq-19:nnudhe}
106(S - 1) = \eta_{\rm He} - \eta_{\rm D} = 5.54 \pm 3.78,
\end{equation}
\begin{equation}
\label{eq-19:etadhe}
106\eta_{10} = 100\eta_{\rm D} + 6\eta_{\rm He} = 665 \pm 36.
\end{equation}
In Fig.\,\ref{fig-19:hevsd} are shown contours of constant values of $\eta_{10}$ and \Deln~in the Y$_{\rm P} - y_{\rm DP}$ plane, along with the observationally inferred values of $y_{\rm DP}$ and \Yp~and their $1\,\sigma$ error bars.  Measurements of $y_{\rm DP}$ and \Yp~constrain $\eta_{10}$ and \Deln.  From BBN, using D and \4he, it is found that $\eta_{10} = 6.27 \pm 0.34$ ($\Omega_{\rm B}h^{2} = 0.0229 \pm 0.0012$) and $\Delta{\rm N}_{\nu} = 0.66^{+0.47}_{-0.45}$ (N$_{eff} = 3.71^{+0.47}_{-0.45}$).  Fig.\,\ref{fig-19:nnuvseta95} shows the 68\% and 95\% confidence contours corresponding to these results.  While $\Delta{\rm N}_{\nu} \approx 1$ (a sterile neutrino?) is somewhat favored, this result is also consistent with no dark radiation (\Deln~= 0) within 95\% confidence.  However, as may be seen from Fig.\,\ref{fig-19:nnuvseta95}, the presence of two sterile neutrinos is disfavored at $\ga 95\%$ confidence. 

The observationally-inferred values of $\eta_{\rm D}$ and $\eta_{\rm He}$ may be used to predict the primordial abundance of lithium synthesized during BBN.  
\begin{equation}
\label{litetannu}
106\eta_{\rm Li} = 103\eta_{\rm D} + 3\eta_{\rm He} = 648 \pm 31.
\end{equation}
This leads to the prediction that A(Li) $= 2.70 \pm 0.06$, far in excess of the lithium abundances inferred from the observations of the most metal-poor stars in the Galaxy, reinforcing the SBBN result of a lithium problem (see \S\,\ref{sec-19:sm}).

\subsection{BBN Constraints On The Baryon Density ($\eta_{10}$) And Lepton Asymmetry ($\xi$)}
\label{sec-19:etaxi}

\begin{figure}[h!!]
\begin{center}
\vspace{1cm}
 \includegraphics*[scale=0.5]{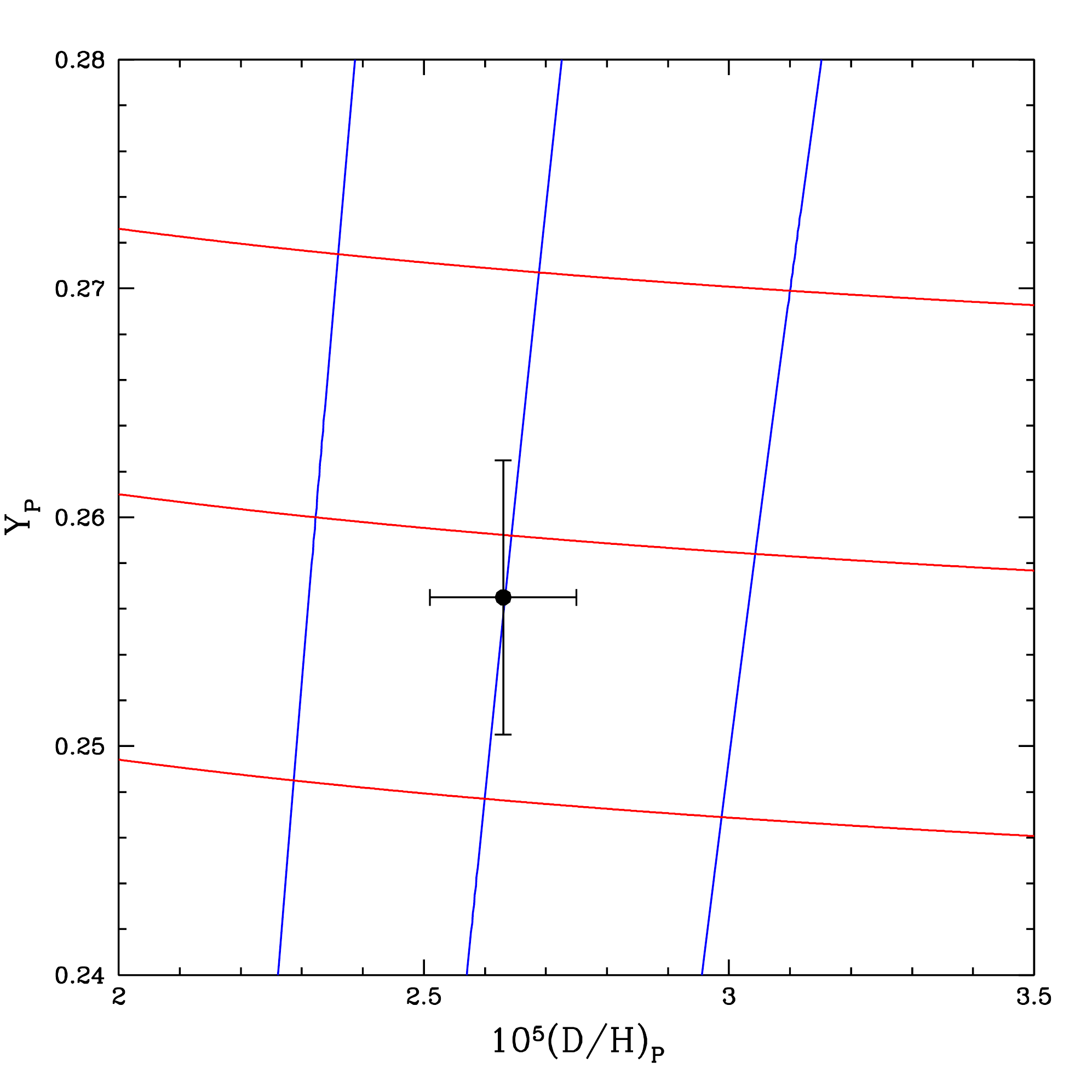}
\caption{Contours of constant values of $\xi$~(red) and $\eta_{10}$ (blue) in the Y$_{\rm P} - y_{\rm DP}$ plane.  From bottom to top the red curves correspond to $\Delta{\rm N}_{\nu} = 0, -0.05, -0.10$.  From left to right the blue curves correspond to $\eta_{10} = 6.5, 6.0, 5.5$.  Also shown (filled circle and error bars) are the adopted primordial abundances of D and \4he and their $1\,\sigma$ uncertainties.
  \label{fig-19:hevsdxi}}
\end{center}
\end{figure}

\begin{figure}[h!!]
\begin{center}
\vspace{1cm}
 \includegraphics*[scale=0.5]{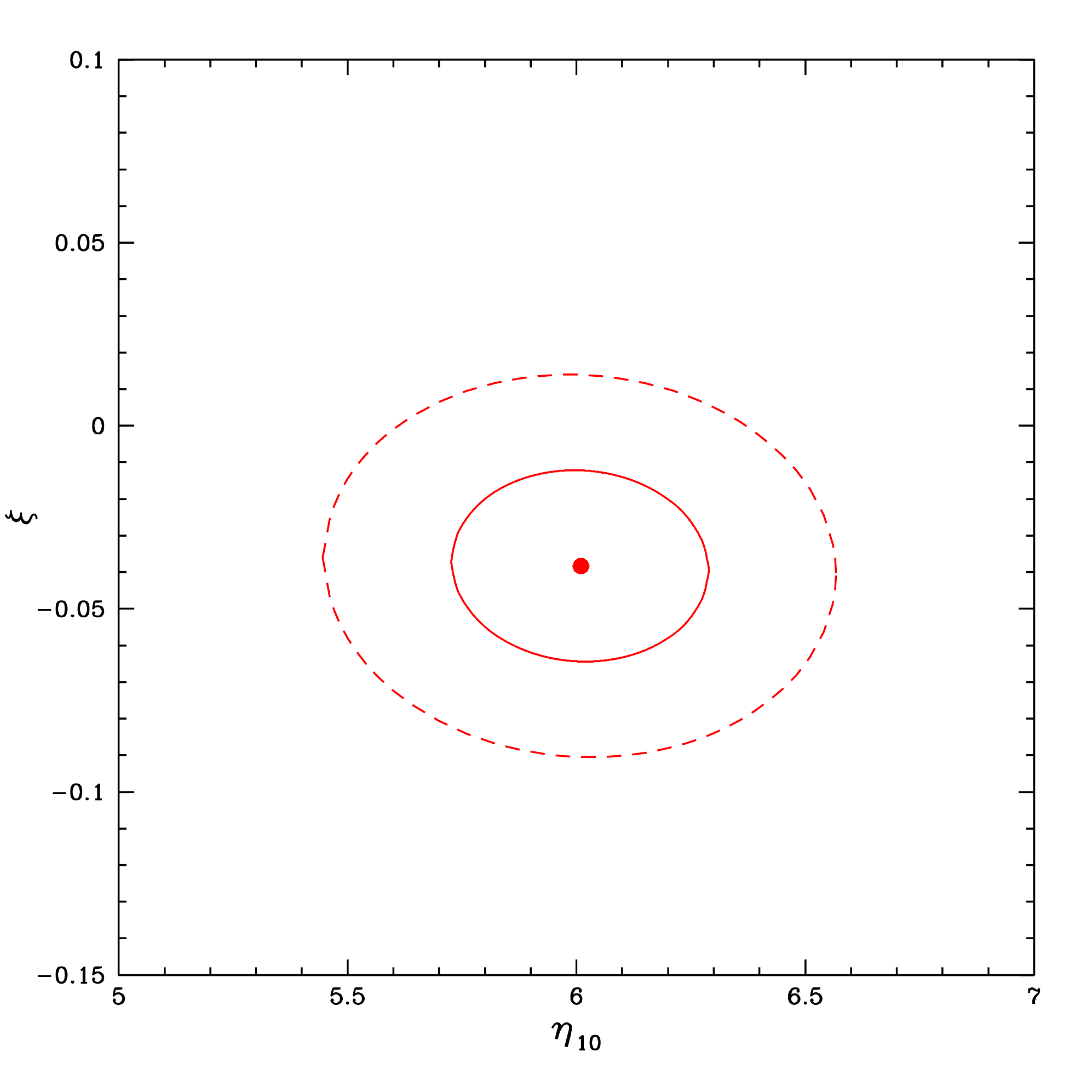}
\caption{The BBN-inferred 68\% (solid) and 95\% (dashed) contours in the $\xi - \eta_{10}$~plane derived from D and \4he assuming that \Deln~= 0.
\label{fig-19:xivseta95}}
\end{center}
\end{figure}

If it is assumed that there is no dark radiation (\Deln~= 0), the observationally-inferred abundances of D and \4he may be used to constrain the baryon density and any lepton asymmetry.
\begin{equation}
\label{eq-19:xidhe}
145\xi = \eta_{\rm D} - \eta_{\rm He} = -5.54 \pm 3.78,
\end{equation}
\begin{equation}
\label{eq-19:xietadhe}
116\eta_{10} = 115\eta_{\rm D} + \eta_{\rm He} = 697 \pm 32.
\end{equation}
In Fig.\,\ref{fig-19:hevsdxi} are shown contours of constant values of $\eta_{10}$ and $\xi$ in the Y$_{\rm P} - y_{\rm DP}$ plane, along with the adopted values of $y_{\rm DP}$ and \Yp~and their $1\,\sigma$ error bars.  Measurements of $y_{\rm DP}$ and \Yp~constrain $\eta_{10}$ and $\xi$.  From BBN using the adopted primordial D and \4he~abundances it is found in this case that $\eta_{10} = 6.01 \pm 0.28$ ($\Omega_{\rm B}h^{2} = 0.0219 \pm 0.0010$) and $\xi = -0.038 \pm 0.026$.  The latter result is consistent with $\xi = 0$ at $\sim 1.5\,\sigma$.  At $2\,\sigma$, this result provides an upper bound to the {\it magnitude} of the neutrino degeneracy parameter ($|\xi| \la 0.090$) which can be used to constrain the contribution to $\Delta{\rm N}_{\nu}$ resulting from the presence of the ``extra" energy density associated with an excess of neutrinos over antineutrinos or, vice-versa: $\Delta{\rm N}_{\nu}(\xi) \la 0.011$ at $\sim 2\,\sigma$.  Fig.\,\ref{fig-19:xivseta95} shows the 68\% and 95\% confidence contours corresponding to these results.  This result is consistent with no lepton asymmetry to better than 95\% confidence.

As before (see \S\,\ref{sec-19:etannu}) the observationally-inferred values of $\eta_{\rm D}$ and $\eta_{\rm He}$ may be used to predict the BBN lithium abundance

\begin{equation}
\label{litetaxi}
145\eta_{\rm Li} = 142\eta_{\rm D} + 3\eta_{\rm He} = 881 \pm 41.
\end{equation}
This leads to the prediction that A(Li) $= 2.69 \pm 0.06$, in almost exact agreement with the corresponding prediction in the presence of dark radiation (assuming that $\xi = 0$), reinforcing, once again, the lithium problem.

\section{Discussion}

The simplest (least interesting?) assumption is that of no new physics; the standard model with no dark radiation (\eg~no sterile neutrinos) or a significant lepton asymmetry ($\eta_{\rm B} \ll \eta_{\rm L} \ll 1$).  In this case the SBBN-predicted baryon abundance (see \S\,\ref{sec-19:sm}), $\Omega_{\rm B}h^{2} = 0.0218 \pm 0.0010$, is in excellent agreement with the more precise value found from the CMB, $\Omega_{\rm B}h^{2} = 0.0226 \pm 0.0004$ \cite{19-Komatsu:2011}.  For this value of the baryon density (and \Deln~$= \xi = 0$) the SBBN-predicted abundance of \3he is consistent with the primordial value inferred from Galactic observations and, the relic abundance of \4he agrees with the observationally-inferred value adopted here, within $\sim 1.5\,\sigma$.  As is by now well established, the predicted relic lithium abundance exceeds the values of the lithium abundance inferred from observations of the most metal-poor stars in the Galaxy by factor of $\sim 3 - 4$.

Setting aside for the moment the possibility of a large neutrino degeneracy ($\eta_{\nu} \gg \eta_{\rm B}$) but, allowing for the presence of dark radiation (\Deln~$\neq 0$), BBN along with the adopted primordial abundances of D and \4he may be used to constrain \Deln~and $\Omega_{\rm B}h^{2}$.  In this case (see \S\,\ref{sec-19:etannu}) it is found that $\Delta{\rm N}_{\nu} = 0.66^{+0.47}_{-0.45}~({\rm N}_{eff} = 3.71^{+0.47}_{-0.45}$) and $\Omega_{\rm B}h^{2} = 0.0229 \pm 0.0012$, in excellent agreement with the values of these parameters inferred from various CMB observations \cite{19-Komatsu:2011,19-ACT:2011,19-SPT:2011,19-SPT:2011a}.  When derived from the CMB, the errors on N$_{eff} = 3.046 + \Delta{\rm N}_{\nu}$ are larger (typically by a factor of $\sim 1.5 - 2$), and those on $\Omega_{\rm B}h^{2}$ smaller (typically by a factor of $\sim 3$), than the corresponding BBN uncertainties.  This is likely to change when the PLANCK collaboration analyzes its CMB data.  The PLANCK constraint on \Deln~is expected to be more precise compared to the BBN value by a factor of $\sim 2.5$, while that on the baryon density parameter should be more precise than the BBN value by an order of magnitude \cite{19-Hamann:2008,19-Galli:2010}.  The BBN-predicted relic lithium abundance when $\Delta{\rm N}_{\nu} \neq 0$ is hardly changed from the SBBN case, reinforcing the lithium problem.  As may be seen from Fig.\,\ref{fig-19:nnuvseta95}, while the result for \Deln~(N$_{eff}$) is closer to \Deln~= 1 than to \Deln~= 0, offering some support for the existence of one sterile neutrino, it is consistent with \Deln~= 0 at $\sim 1.5\,\sigma$.  In contrast, the existence of two sterile neutrinos is disfavored by the BBN data at $\ga 95\%$ confidence.  

\begin{figure}[h!!]
\begin{center}
\vspace{1cm}
 \includegraphics*[scale=0.5]{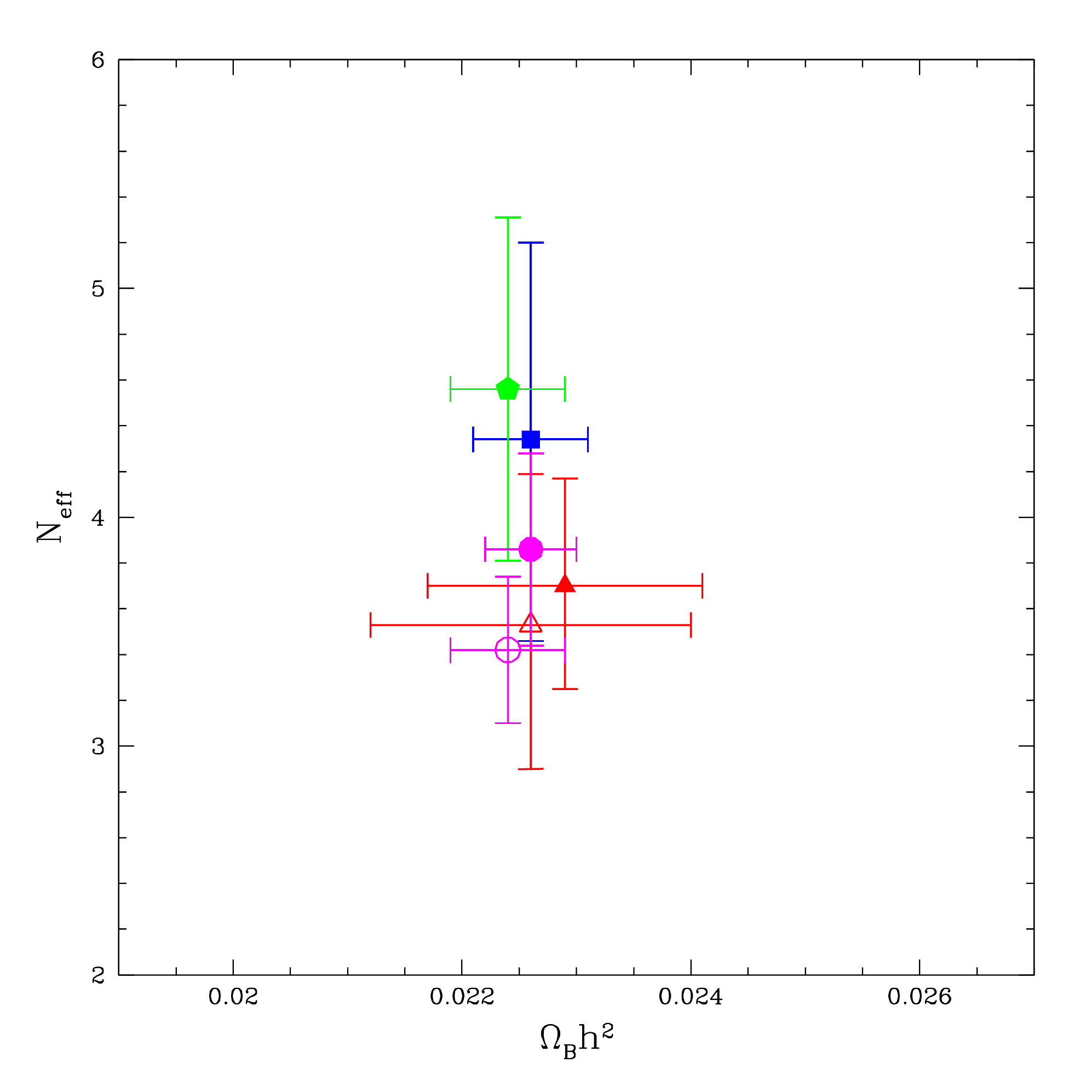}
\caption{Comparing the BBN predictions of N$_{eff}$ and $\Omega_{\rm B}h^{2}$ with those from various CMB determinations: BBN D + \4he (red filled triangle), BBN D + WMAP7 \cite{19-Komatsu:2011} $\Omega_{\rm B}h^{2}$ (red open triangle), WMAP7 \cite{19-Komatsu:2011} (blue filled square), ACT \cite{19-ACT:2011} (green filled pentagon), SPT \cite{19-SPT:2011} (purple filled circle), SPT + Clusters \cite{19-SPT:2011a} (purple open circle).
\label{fig-19:neffvsomeg}}
\end{center}
\end{figure}

In Fig.\,\ref{fig-19:neffvsomeg} the BBN constraints on \Deln~and $\Omega_{\rm B}h^{2}$ are compared with those from recent CMB analyses.  Here, too, it appears that current data have a preference for one sterile neutrino while being slightly inconsistent with two sterile neutrinos (\eg~BBN and SPT in Fig.\,\ref{fig-19:neffvsomeg}) or with no dark radiation (\eg~WMAP7 and ACT in Fig.\,\ref{fig-19:neffvsomeg}).

Unlike the CMB, BBN has the potential to probe a non-zero (albeit relatively large) lepton asymmetry (neutrino degeneracy).  Current data, driven by the adopted \4he abundance, are consistent with a small, negative value for the neutrino degeneracy parameter, $\xi = -0.038 \pm 0.026$ which, however, is only $\sim 1.5\,\sigma$ from zero.  A more precise result will only come when (if) there is a reduction in the error of the observationally-inferred value of \Yp, whose uncertainty is dominated by systematics.

\subsection{Sensitivity Of \Deln~And $\xi$ To Primordial Helium And Its Uncertainty}
\label{sec-19:hevstime}

\begin{figure}[h!!]
\begin{center}
\vspace{1cm}
 \includegraphics*[scale=0.4]{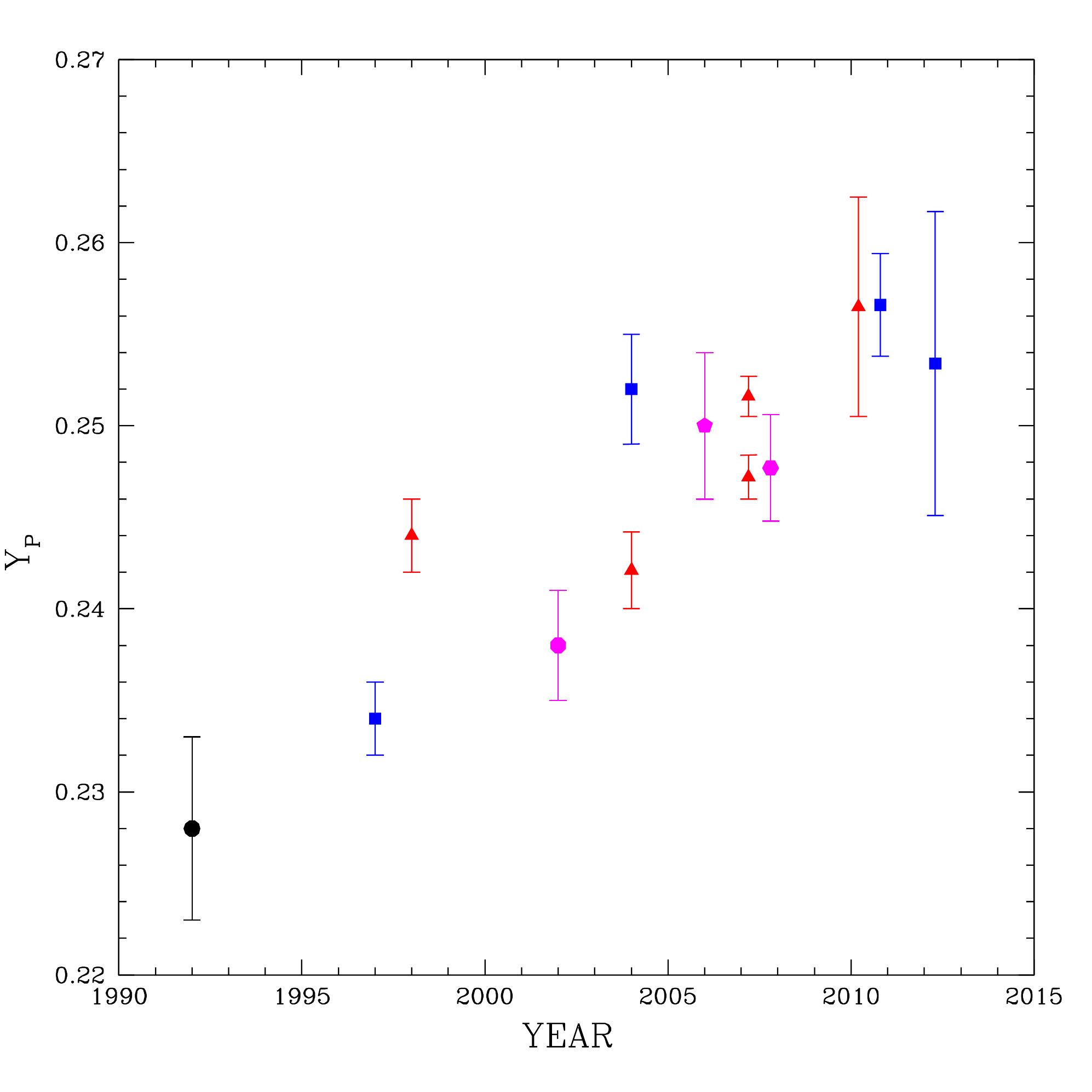}
\hspace{.5cm}
 \includegraphics*[scale=0.4]{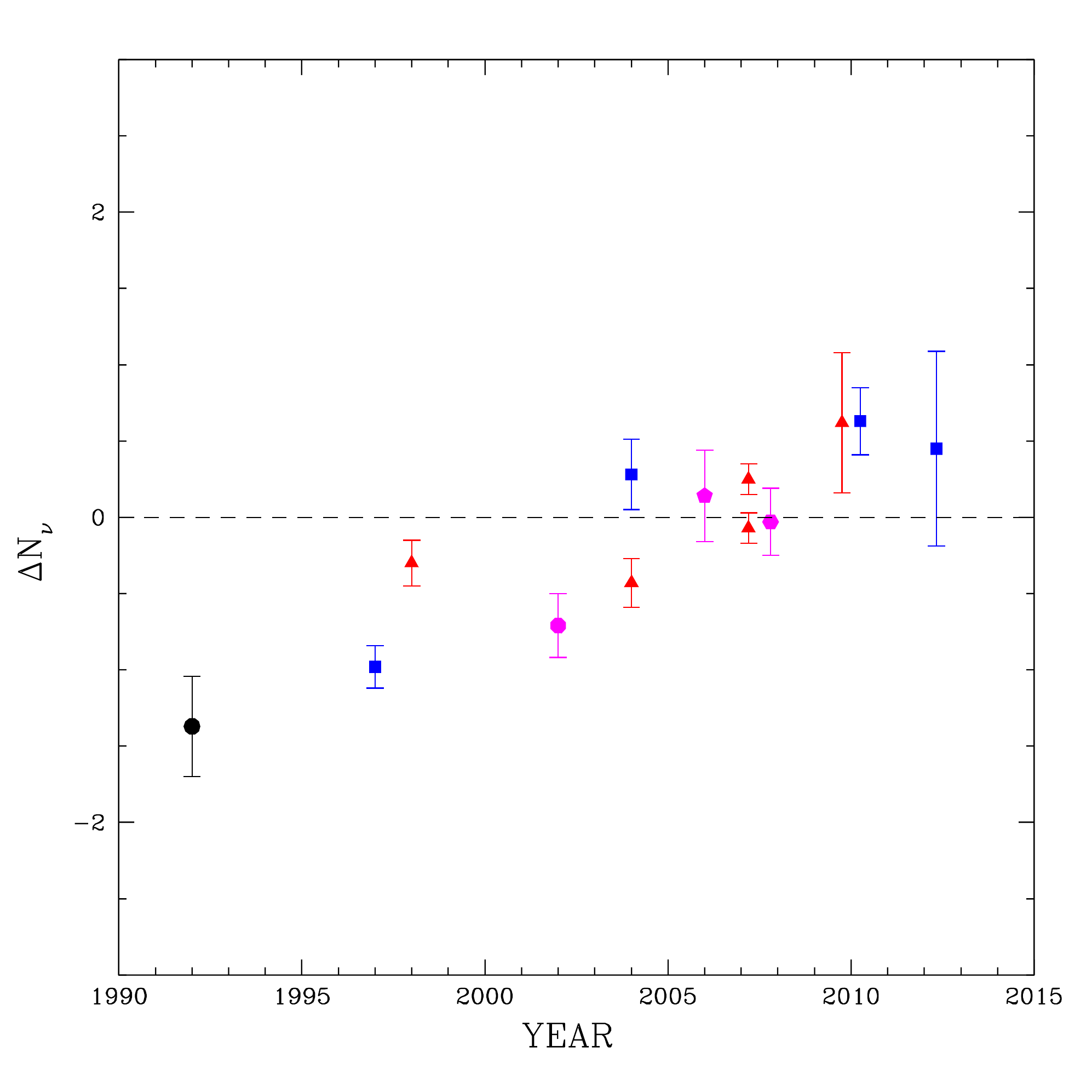}
\hspace{.5cm}
 \caption{The left panel shows a history of the primordial helium mass fraction (\Yp) determinations as a function of time.  The same symbols/colors correspond to determinations from collaborations involving many of the same participants and/or the same observational data.  The right panel shows the corresponding chronology of BBN-determined values of \Deln.  The dashed line shows the SM result, \Deln~= 0.
  \label{fig-19:hevstime}}
\end{center}
\end{figure}

The BBN-predicted helium abundance is sensitive to the early Universe expansion rate ($S$ or, dark radiation \Deln) and to a lepton asymmetry ($\xi$) and very insensitive to the baryon density ($\eta_{\rm B}$).  The results which have been presented here for \Deln~and $\xi$ are mainly driven by the adopted value for the primordial helium abundance, and its uncertainty.  But, the observationally inferred helium abundance, \Yp, is a quantity which has changed dramatically over time as more and better data have been acquired and more careful analyses of the data have been performed.  In the left panel of Fig.\,\ref{fig-19:hevstime} is shown a chronology, over the past $\sim 20$ years, of the published observational determinations of the primordial helium mass fraction, revealing a nearly monotonic increase of \Yp~with time.  In the right panel of Fig.\,\ref{fig-19:hevstime} the chronology of the corresponding \Deln~values is shown, mirroring the increase in \Yp.  Notice that only very recently, within the past 5 -- 7 years, do the data begin to favor $\Delta{\rm N}_{\nu} > 0$.

\subsection{Constraints On \Deln~From BBN D And The CMB-Inferred Baryon Density}
\label{sec-19:bbncmb}

The extreme sensitivity of the BBN-inferred estimates of \Deln~to the adopted helium abundance (and its large errors), is responsible for the relatively large error in the BBN-inferred value of \Deln.  An alternate approach avoiding \4he has been suggested by Nollett and Holder (2011) \cite{19-Nollett:2011} (see, also, Pettini \& Cooke (2012) \cite{19-Pettini:2012}).  

\begin{figure}[h!!]
\begin{center}
\vspace{0.75cm}
 \includegraphics*[scale=0.5]{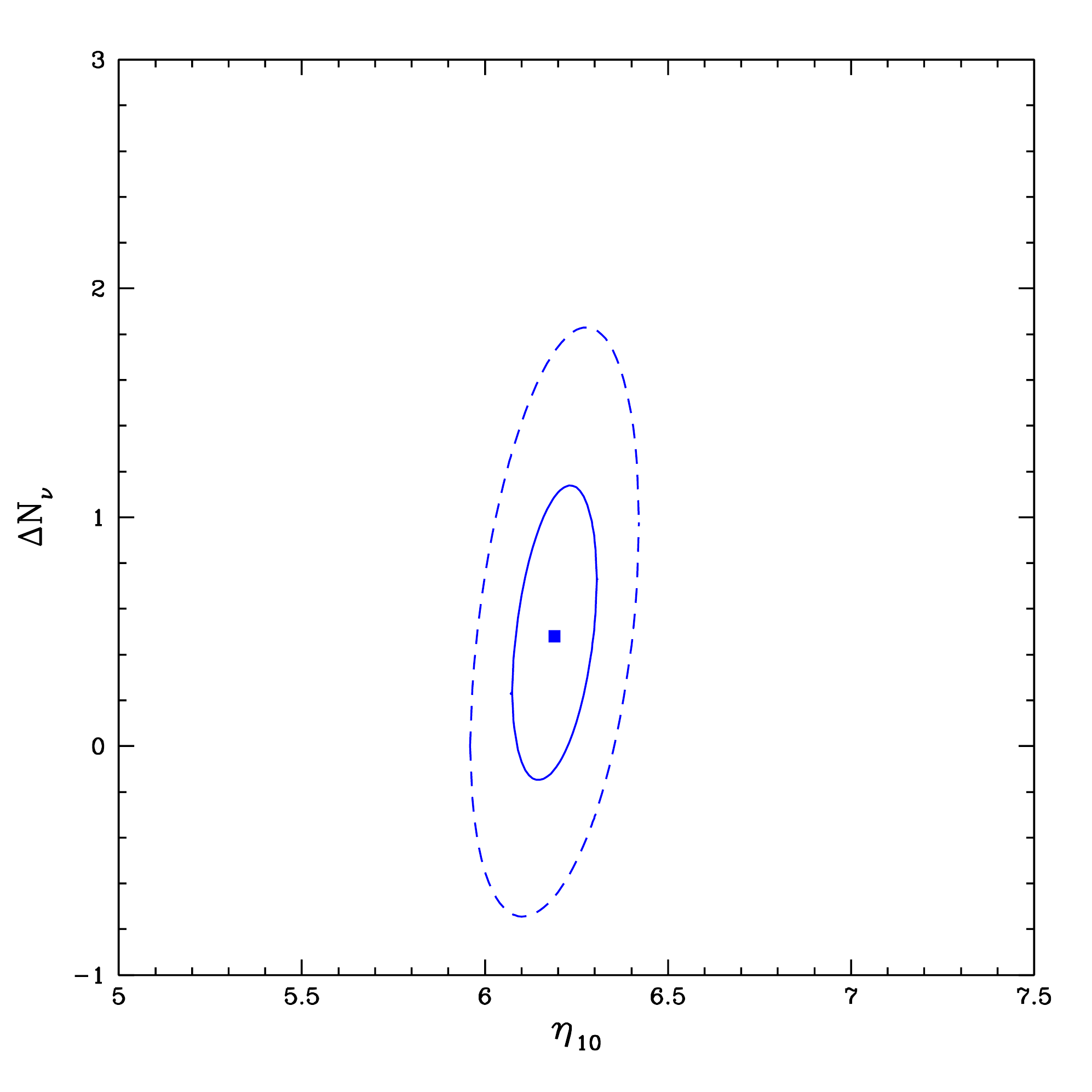}
\caption{The 68\% (solid) and 95\% (dashed) contours in the \Deln~-- $\eta_{10}$~plane derived from BBN deuterium and the CMB constraint on the baryon density \cite{19-Komatsu:2011}.
\label{fig-19:cmb95}}
\end{center}
\end{figure}

In the best of all worlds the BBN-inferred parameter values should be compared with those inferred, independently, from the CMB, complemented when necessary to break degeneracies among the parameters by other astrophysical data from, \eg~large scale structure, supernovae, and the Hubble constant.  In the presence of possible new physics, this would enable a probe of the constancy (or not) of these parameters in the early Universe epochs from BBN until recombination.  However, if it is {\it assumed} that \Deln~and $\eta_{10}$ are unchanged from BBN to recombination, the information provided by BBN using the helium abundance may be replaced with that from the CMB-determined baryon density: $\eta_{10}({\rm CMB}) = 6.190 \pm 0.115$ \cite{19-Komatsu:2011}.  Using this value in combination with deuterium, $\eta_{\rm D} = \eta_{10} - 6(S - 1) = 5.96 \pm 0.28$ (for $\xi = 0$), leads to a smaller estimate of \Deln~but, with a larger uncertainty resulting from the much weaker dependence of $\eta_{\rm D}$ on \Deln: $\Delta{\rm N}_{\nu} = 0.48^{+0.66}_{-0.63}$ (N$_{eff} = 3.53^{+0.66}_{-0.63}$)\footnote{For the single, most precise deuterium abundance found by Pettini \& Cooke \cite{19-Pettini:2012}, $\eta_{\rm D} = 6.10 \pm 0.24$.  If this abundance is identified with the primordial deuterium abundance, $\Delta{\rm N}_{\nu} = 0.18 \pm 0.55$ (N$_{eff} = 3.22 \pm 0.55$).}.  The 68\% and 95\% contours for these results are shown in Fig.\,\ref{fig-19:cmb95}.  While this provides some support, once again, for the presence of one sterile neutrino, the absence of dark radiation ($\Delta{\rm N}_{\nu} = 0$) is consistent with these results at the $\sim 68\%$ confidence level.  For these values of $\eta_{\rm D}$ and $\eta_{10}$ the BBN-predicted helium abundance is Y$_{\rm P} = 0.2541 \pm 0.0081$, very close to(well within the errors of) the observationally-inferred value adopted here.  Once again, the BBN-predicted lithium abundance is a problem: A(Li) $= 2.69 \pm 0.05$.  

This approach, replacing \4he with the CMB determined baryon density parameter, could also be used to constrain a lepton asymmetry.  The corresponding constraint on the neutrino degeneracy, $\xi = -0.18 \pm 0.24$, while entirely consistent with $\xi = 0$, has an uninterestingly large uncertainty resulting from the very weak dependence of $\eta_{\rm D}$ on $\xi$.

\subsection{Supplementing BBN With The CMB To Constrain $\xi \neq 0$ And $\Delta{\rm N}_{\nu} \neq 0$}
\label{sec-19:xunnu}

\begin{figure}[h!!]
\begin{center}
\vspace{1cm}
 \includegraphics*[scale=0.5]{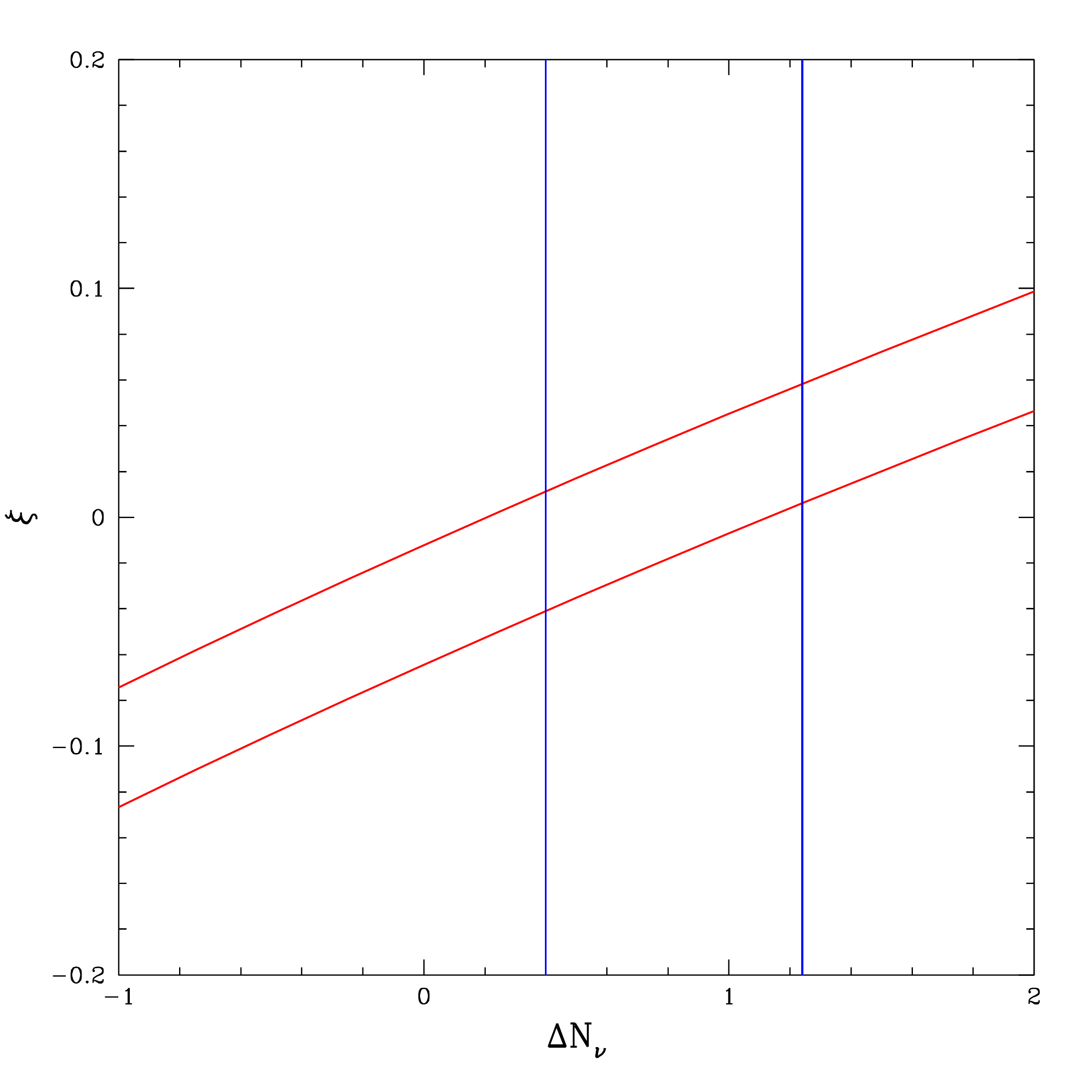}
\caption{The $\pm 1\,\sigma$ band (red) in the $\xi - \Delta{\rm N}_{\nu}$ plane from BBN using the D and \4he constraints.  The blue band is the $\pm 1\,\sigma$ range for \Deln~from Joudaki (2012)~\cite{19-Joudaki:2012}.
\label{fig-19:xivsnnu}}
\end{center}
\end{figure}

The BBN-predicted primordial light element abundances depend on all three of the key parameters \{$\eta_{10}, \Delta{\rm N}_{\nu}, \xi$\}.  However, the uncertainty in the observationally-inferred relic abundance of \3he, along with the lithium problem(s), leaves only two, relatively well constrained primordial abundances, those for D and \4he.  From BBN alone and these abundances, all three parameters can't be determined independently but, one of them can be eliminated resulting in a relation (degeneracy) between the remaining two.  For example, $\eta_{\rm D}$ and $\eta_{\rm He}$ may be used to eliminate $\eta_{10}$, leading to $\xi = \xi(\Delta{\rm N}_{\nu};\,y_{\rm DP},{\rm Y_{\rm P}})$, where
\begin{equation}
145\xi = 106(S - 1) + \eta_{\rm D} - \eta_{\rm He}.
\label{eq-19:xivsnnudhe}
\end{equation}
This constraint on $\xi$ versus \Deln~is shown by the red band in Fig.\,\ref{fig-19:xivsnnu}.  Without an independent constraint on \Deln, the degeneracy between $\xi$ and \Deln~seen in Fig.\,\ref{fig-19:xivsnnu} cannot be broken.  The CMB provides such a constraint.  If BBN, using the observationally-inferred D and \4he abundances, is now supplemented with an independent constraint on \Deln~from the CMB, then a combined constraint, allowing for both dark radiation and lepton asymmetry, may be found.  Using the CMB results from WMAP7 \cite{19-Komatsu:2011}, ACT \cite{19-ACT:2011}, and the SPT \cite{19-SPT:2011,19-SPT:2011a}, along with complementary constraints from large scale structure, the Hubble constant, supernovae, and galaxy clusters, Joudaki (2012) \cite{19-Joudaki:2012} finds N$_{eff} = 3.87 \pm 0.42$, corresponding to $S = 1.065 \pm 0.032$.  Using this result in Eq.\,\ref{eq-19:xivsnnudhe} leads to $\xi = 0.009 \pm 0.035$, entirely consistent with $\xi = 0$.  The $2\,\sigma$ upper bound to $|\xi|$ is 0.079, corresponding to an upper bound to $\Delta{\rm N}_{\nu}(\xi) = 0.008$.  Once again, the BBN-predicted lithium abundance is high, A(Li)~$= 2.70 \pm 0.06$, reinforcing the lithium problem.

A similar approach may be used to eliminate \Deln~instead of $\eta_{10}$ to find $\xi = \xi(\eta_{10};\,y_{\rm DP},{\rm Y_{\rm P}})$ and, to use the CMB for a constraint on $\eta_{10}$.  But this approach, which is also consistent with $\xi = 0$, leads to a less precise constraint: $\xi = -0.012 \pm 0.052$.

\section{Summary And Anticipation Of Future Results}
\label{sec-19:summary}

The BBN results presented here provide modest support for the presence of dark radiation ($\Delta{\rm N}_{\nu} \ga 0$ at $\sim 1.5\,\sigma$), to go along with the more robust evidence for dark matter and dark energy.  The current observational data (D and \4he), while allowing for the presence of a sterile neutrino (\Deln~= 1 at $\la 1\,\sigma$), disfavors the presence of two sterile neutrinos ($\Delta{\rm N}_{\nu} = 2$ at $\sim 2.8\,\sigma$).  If, instead, it is {\it assumed} that \Deln~= 0, a small, but non-zero lepton asymmetry is favored, also at the $\sim 1.5\,\sigma$ confidence level.  In contrast, if both \Deln~and $\xi$ are allowed to vary freely and, if BBN (D and \4he) is supplemented by a CMB constraint on \Deln, a vanishing lepton asymmetry ($\xi = 0.009 \pm 0.035$) is favored.

Currently, as may be seen from Fig.\,\ref{fig-19:neffvsomeg}, there is very good agreement between the BBN and CMB constraints on the baryon density and dark radiation (the CMB is insensitive to a small or modest lepton asymmetry).  For many years BBN provided the best constraints on the baryon density ($\eta_{10}~{\rm or}~\Omega_{\rm B}h^{2}$) and on dark radiation (\Deln~or $S$), as well as the only constraint on lepton asymmetry.  With WMAP7 \cite{19-Komatsu:2011} and other CMB datasets \cite{19-ACT:2011,19-SPT:2011,19-SPT:2011a} the best constraints on the baryon density now are from the CMB, which allow for a factor of $\sim 2 - 3$ more precise determination of $\eta_{10}$.  However, in the present, pre-PLANCK era, BBN still provides the best dark radiation constraint, albeit with an uncertainty smaller than that from the CMB by only a factor of $\sim 1.5 - 2$.  It is expected that with the publication of the PLANCK data the dark radiation torch will pass to the CMB.  Depending on what PLANCK finds, it may be possible to establish the presence of dark radiation (a sterile neutrino?) at the $\sim 5\,\sigma$ level {\it if}, for example, PLANCK should find, \Deln~$= 1 \pm 0.2$.  If, however, PLANCK should find (the best or worst of all worlds?) \Deln~$= 0.5 \pm 0.2$, the presence dark radiation will be favored but, that of a sterile neutrino will be somewhat disfavored\footnote{It is perhaps worth recalling that the contribution of a light, thermalized scalar corresponds to \Deln~= 4/7 = 0.57.}.

For all the possibilities considered here (\Deln~$= \xi = 0$; $\xi = 0$, $\Delta{\rm N}_{\nu} \neq 0$; $\Delta{\rm N}_{\nu} = 0$, $\xi \neq 0$; $\Delta{\rm N}_{\nu} \neq 0$, $\xi \neq 0$), the BBN-predicted lithium abundance hardly changed at all ($2.68 \pm 0.06 \leq {\rm A(Li)} \leq 2.70 \pm 0.06$).  This insensitivity is easy to understand since the BBN-predicted lithium and deuterium abundances are strongly correlated
\begin{equation}
\eta_{\rm Li} = \eta_{\rm D} + 3[(S - 1) - \xi].
\label{eq-19:livsd}
\end{equation}
Because the values of $3(S - 1)$ and $3\xi$ are almost always small compared to $\eta_{\rm D}$, the corrections to a perfect lithium -- deuterium correlation are generally at only the few percent level.  A solution to the lithium problem is not to be found with dark radiation or a lepton asymmetry.

\subsection{Anticipating The Future}

The future for the key parameters related to the baryon abundance ($\eta_{\rm B}$) and the presence, or not, of dark radiation (\Deln) lies with the CMB and the anticipated results from the PLANCK mission. While it is impossible to predict the central values PLANCK will find for $\eta_{\rm B}$ or \Deln, it is possible to forecast the precision to be expected from the PLANCK data and analyses \cite{19-Hamann:2008,19-Galli:2010}.  Such forecasts suggest that the uncertainty in the baryon abundance determination will be of order $\sigma(\eta_{10}) \approx 0.03$, nearly an order of magnitude better than the current BBN precision.  The same forecasts suggest that \Deln~will be constrained to $\sigma($\Deln$) \approx 0.2$, or better.  This would result in an improvement over the current BBN precision by a factor of $\sim 2.5$.  

If it is {\it assumed} that the PLANCK values of the key parameters are identical to those at BBN, ignoring their possible evolution from the epoch of BBN until recombination, then these values may be used in combination with BBN to predict the relic abundances\footnote{In contrast with the current approach of using BBN in concert with the observationally inferred relic abundances to predict the values of the key parameters.}.  For example, for deuterium, it is anticipated that PLANCK will constrain $\eta_{\rm D}$ with a precision of $\sigma(\eta_{\rm D}) \approx 0.1$ or, to $\la 2\%$ for $\eta_{\rm D} \approx 6$.  The largest uncertainty in the BBN-predicted deuterium abundance at present and in this anticipated future arises from uncertain nuclear reaction rates \cite{19-Nollett:2011}.  It can be hoped that this uncertainty may be reduced by new laboratory data, reducing the error in the BBN-predicted value of $y_{\rm DP}$ by perhaps a factor of $\sim 2$.  This would lead to a reduction in the error in the inferred value of $\eta_{\rm D}$ by nearly a factor of two, $\sigma(\eta_{\rm D}) \approx 0.3 \rightarrow  0.15$.  

For helium, PLANCK may constrain $\eta_{\rm He}$ to $\sigma(\eta_{\rm He}) \approx 1.6$ which, while still large, is a factor $\sim 2.3$ smaller than the current BBN uncertainties.  This suggests that using the CMB determined values of $\eta_{10}$ and \Deln, the BBN-predicted primordial helium mass fraction will be known to $\sigma({\rm Y_{P}}) \la 0.003$, a precision anticipated to also be attainable in an independent determination of \Yp~from the CMB \cite{19-Hamann:2008,19-Galli:2010}.

For lithium, PLANCK may constrain $\eta_{\rm Li}$ to $\sigma(\eta_{\rm Li}) \approx 0.06$ or, to better than $\sim 1\%$ for $\eta_{\rm Li} \approx 6$.  However, as for deuterium, the precision of the BBN-predicted primordial lithium abundance is limited by the nuclear physics uncertainties ($\sim 10\%$).  Nonetheless, it will be very interesting to see if the PLANCK data support or, possibly eliminate, the lithium problem(s).

Although the CMB is insensitive to a lepton asymmetry, as may be seen from Eq.\,\ref{eq-19:xivsnnudhe}, a combination of BBN and CMB constraints on \Deln, $y_{\rm DP}$, and \Yp~can constrain a neutrino degeneracy, provided that the lepton asymmetry is very large compared to the baryon asymmetry.  For example, for the anticipated CMB constraints on $\sigma(\Delta{\rm N}_{\nu}) \approx 0.2$ and on the primordial helium abundance, $\sigma({\rm Y_{P}}) \approx 0.003$, along with a BBN constraint on $y_{\rm DP}$, $\sigma_{\xi} \approx 0.018$ or, $\sigma_{\eta_{\rm L}} \approx (\pi^{2}/4\zeta(3))\sigma_{\xi} \approx 0.036 \approx 6\times 10^{7}\eta_{\rm B}$.

\subsection{Summary}

This review finds itself on the cusp of potentially great changes.  Current BBN and CMB data provide strong support for the presence of (at least) three SM neutrinos, thermally populated during the early evolution of the Universe. This provides indirect support for the so far invisible, relic neutrino background.  The new CMB and large scale structure data have the potential to constrain the baryon asymmetry and the presence, or not, of dark radiation to new levels of precision, testing BBN and the current estimates of the relic abundances of the light elements.  It will be of great interest to compare and contrast the current BBN results with those from the new data and to see what we may learn about new physics, including neutrino physics, beyond the standard models of particle physics and cosmology.

\begin{center}
{\bf Acknowledgments}
\end{center}
I am pleased to acknowledge informative conversations and email exchanges with R. Cyburt, M. Fumagalli, K. Nollett, M. Pettini, J.~X. Prochaska.  My research is supported at OSU by the DOE.  

  \end{document}